# Observation of superconducting collective modes from competing pairing instabilities in single-layer $NbSe_2$

*Wen Wan[1,†] Paul Dreher[1,†,] Daniel Muñoz-Segovia[1], Rishav Harsh[1], Haojie Guo[2], Antonio J. Martínez-Galera[2,3], Francisco Guinea[1,4], Fernando de Juan[1,5] and Miguel M. Ugeda[1,5,6,*]*

[1]Donostia International Physics Center, Paseo Manuel de Lardizábal 4, 20018 San Sebastián, Spain

[2]Departamento de Física de la Materia Condensada, Universidad Autónoma de Madrid, Madrid E-28049, Spain

[3]Instituto Nicolás Cabrera, Universidad Autónoma de Madrid, Madrid E-28049, Spain

[4]IMDEA Nanoscience, C/Faraday 9, 28049 Madrid, Spain.

[5]Ikerbasque, Basque Foundation for Science, 48013 Bilbao, Spain.

[6]Centro de Física de Materiales, Paseo Manuel de Lardizábal 5, 20018 San Sebastián, Spain.

*† These authors contributed equally to this work.*

** Corresponding author: mmugeda@dipc.org*

**In certain unconventional superconductors with sizable electronic correlations, the availability of closely competing pairing channels leads to characteristic soft collective fluctuations of the order parameters, which leave fingerprints in many observables and allow to scrutinize the phase competition. Superconducting layered materials, where electron-electron interactions are enhanced with decreasing thickness, are promising candidates to display these correlation effects. In this work, we report the existence of a soft collective mode in single-layer $NbSe_2$, observed as a characteristic resonance excitation in high resolution tunneling spectra. This resonance is observed along with higher harmonics, its frequency $\Omega/2\Delta$ is anticorrelated with the local superconducting gap $\Delta$, and its amplitude gradually vanishes by increasing the temperature and upon applying a magnetic field up to the critical values ($T_C$ and $H_{C2}$), which sets an unambiguous link to the superconducting state. Aided by a microscopic model that captures the main experimental observations, we interpret this resonance as a collective Leggett mode that represents the fluctuation towards a proximate *f*-wave triplet state, due to subleading attraction in the triplet channel. Our findings demonstrate the fundamental role of correlations in superconducting 2D transition metal dichalcogenides, opening a path towards unconventional superconductivity in simple, scalable and transferable 2D superconductors.**

## I. Introduction

In the Migdal-Eliashberg theory[1,2] of superconductivity, electron-phonon coupling is responsible for the attraction that binds Cooper pairs together in the standard s-wave channel. In superconductors with significant electronic correlations, however, the Coulomb repulsion can be detrimental for pairing and other mechanisms need to be invoked to explain the emergence of "unconventional" superconductivity, which often occurs with different pairing symmetry in the spin or orbital sectors. Several classes of correlated electron systems such as the cuprates[3–10], iron-pnictides[11–16], iron-chalcogenides[17–19], and several heavy-fermion compounds[20,21], have been identified as unconventional superconductors and more recently, superconductivity with unconventional features has also been identified in twisted bilayer graphene (TBLG)[22]. The different experimental complexities to produce and probe these materials make the study of their unconventional superconductivity a formidable problem.

In this arena, transition metal dichalcogenides (TMD) are promising candidates to provide an alternative route to unconventional superconductivity. Electronic correlations are intrinsically present in this family of layered materials, which manifest in various ways such as in Mott phases[23], magnetic order[24–26], charge/spin density waves[27], quantum spin liquids[28] and superconductivity[22,27,29–34]. Furthermore, in two dimensions, electron–electron interactions are markedly enhanced due to reduced screening, which can enable non-phononic Cooper pairing mechanisms. Unlike all the unconventional superconductors known so far, including TBLG, TMDs can be easily obtained by several methods (molecular beam epitaxy, carbon vapor deposition, atomic layer deposition, exfoliation, etc.), to yield scalable 2D superconductors of simple handling and transfer.

Among 2D TMD superconductors, single-layer $NbSe_2$ has received the most attention and its superconducting properties have been extensively studied[33,34]. Monolayer $NbSe_2$ has a non-centrosymmetric crystal structure (point group $D_{3h}$) which enables a form of spin orbit coupling (SOC) where spins lock out of the plane, leading to Ising superconductivity[31] with enhanced robustness to in-plane magnetic fields[32]. The absence of inversion also enables singlet-triplet mixing[25,35], so far of unknown magnitude. More recently, transport experiments have revealed a two-fold anisotropy of the superconducting state under in-plane magnetic fields, which has been interpreted in terms of a competing nematic superconducting instability[36,37]. In parallel, tunneling junction experiments also claimed the existence of a subleading triplet order parameter to explain the dependence of the gap to in-plane fields in the thin film limit[38]. These experiments suggest sizable electronic correlations as the origin of the competing pairing instabilities. In this work, by means of high-resolution scanning tunneling microscopy and spectroscopy (STM/STS) measurements at 340 mK, we have observed a collective mode univocally associated to superconductivity, which we attribute to a related competing triplet (*f*-wave) pairing channel. This finding strongly suggests that many-body correlations play a dominant role in the emergence of superconductivity among TMD superconductors.

## II. Results

We investigate the superconducting properties of single-layer $NbSe_2$ with samples grown by molecular beam epitaxy on bilayer graphene on SiC(0001) and h-BN/Ir(111) substrates (see SM for details). Since the phenomenology is very similar on both substrates, in the following we will focus on the experiments on $NbSe_2$/graphene (see SM for data on h-BN). **Figure 1**a illustrates the typical morphology of our $NbSe_2$ monolayers on graphene. At low temperatures, single-layer $NbSe_2$ exhibits charge density wave (CDW) order and superconductivity with critical temperatures of $T_{CDW} \approx 33$ K and $T_C \approx 2$ K, respectively[27,30]. Both electronic phases develop a gap feature in the density of states (DOS) at the Fermi level ($E_F$) that can be measured via low-bias STS measurements. The CDW gap in the dI/dV spectra (Figure 1b) appears as a V-shaped dip at $E_F$ bound by coherence peaks with average locations around ± 3-5 mV (ref.[27]). The CDW only gaps out a fraction of the Fermi surface, which allows the development of superconductivity at lower temperatures.

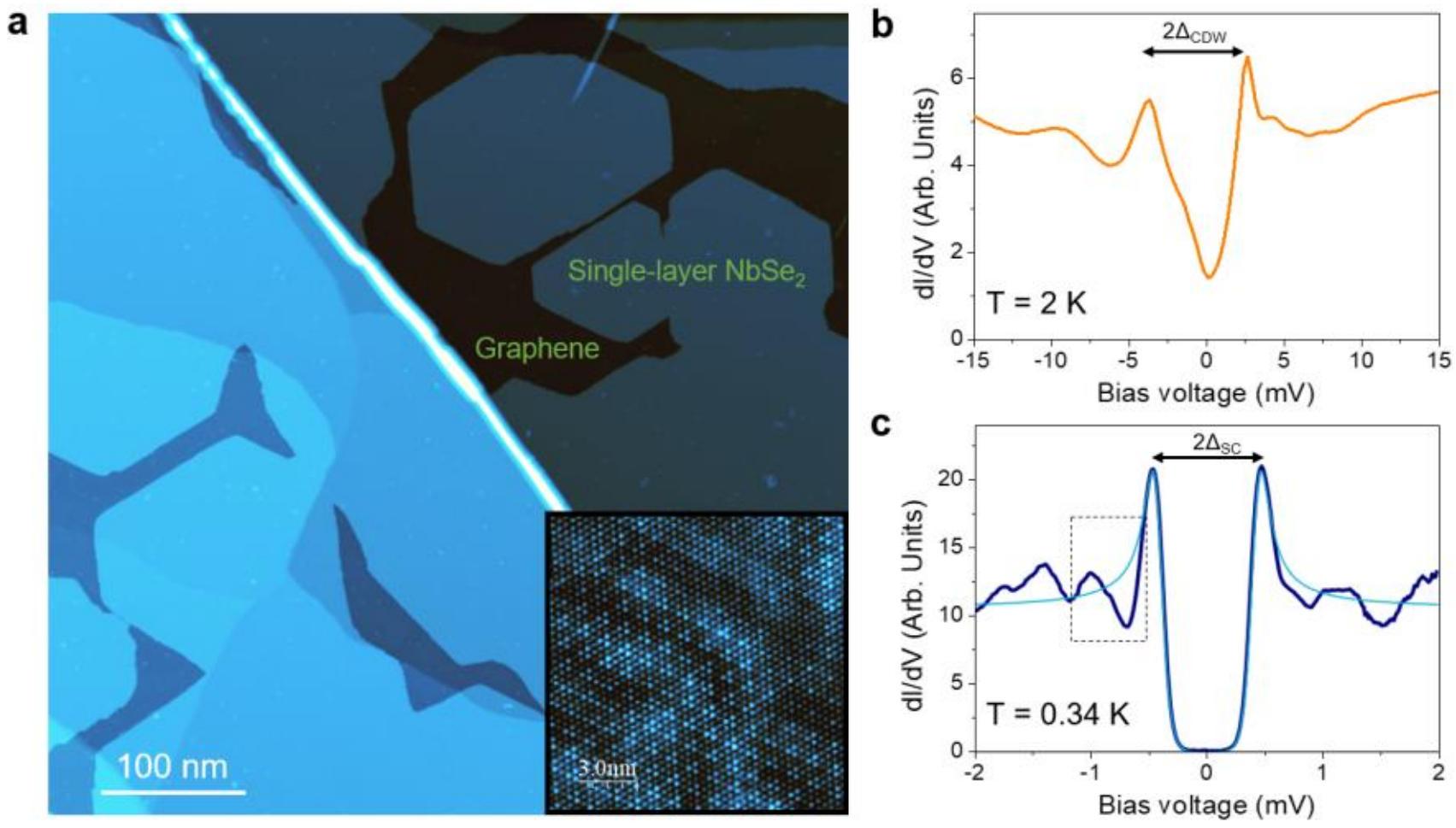

**Figure 1 | Low-energy electronic structure of single-layer $NbSe_2$. a**, Large-scale STM image of single-layer $NbSe_2$/BLG/SiC(0001) in the submonolayer coverage range ($V_s$ = 105 mV, I = 0.01 nA, T = 0.34 K). The inset shows an atomically resolved STM image of the $NbSe_2$ layer showing the 3x3 CDW order ($V_s$ = 30 mV, I = 0.46 nA, T = 0.34 K). **b**,**c**, Low-bias STM dI/dV spectra acquired on single-layer $NbSe_2$ showing the CDW gap ($\Delta_{CDW}$) (f = 833 Hz, $V_{a.c.}$ = 200 μV) in **b** and the superconducting gap (Δ) (f = 833 Hz, $V_{a.c..}$ = 20 μV) in **c**. The boxed region in **c** shows one of the characteristic dip-hump features found in this system.

The fingerprint of the superconducting state in single-layer $NbSe_2$ is shown in Fig.1c, which displays a typical dI/dV curve acquired at T = 0.34 K. This spectrum reveals an absolute gap that fits the BCS gap of width $\Delta_{BCS} = 0.38$ meV (light blue curve). The averaged BCS gap over different locations is $\overline{\Delta_{BCS}} = 0.4\ meV$. As can be seen, however, the experimental conductance for energies higher than the coherence peaks ( $|V| > 0.6\ meV$ ) departs from the BCS conductance and shows several dip-hump satellite features at both polarities, such as those shown in the dashed rectangle in Figure 1c. We note that these STS features are unique to single-layer $NbSe_2$, and are not present in bulk (see SM for STS in bulk $NbSe_2$).

To better describe these spectral features, **Figure 2**a shows four dI/dV curves taken in different locations. These curves reveal the existence of multiple dip-hump features (or peaks, see SM) at both polarities, which are seen usually symmetric with respect to $E_F$ and nearly equidistant. We performed statistical analysis over 2855 dI/dV curves acquired at T = 0.34 K in different spatial locations, using several samples and tips (see SM). As seen in the histogram of Figure 2b, this analysis reveals the existence of three clear satellite peaks within $|V| = 3\ mV$ (both polarities exhibit similar statistics). A much weaker and wider fourth peak is also present in the histogram, but since its energy is already close to the CDW coherence peaks, we do not believe it to be a replica and, therefore, we do not consider it further.

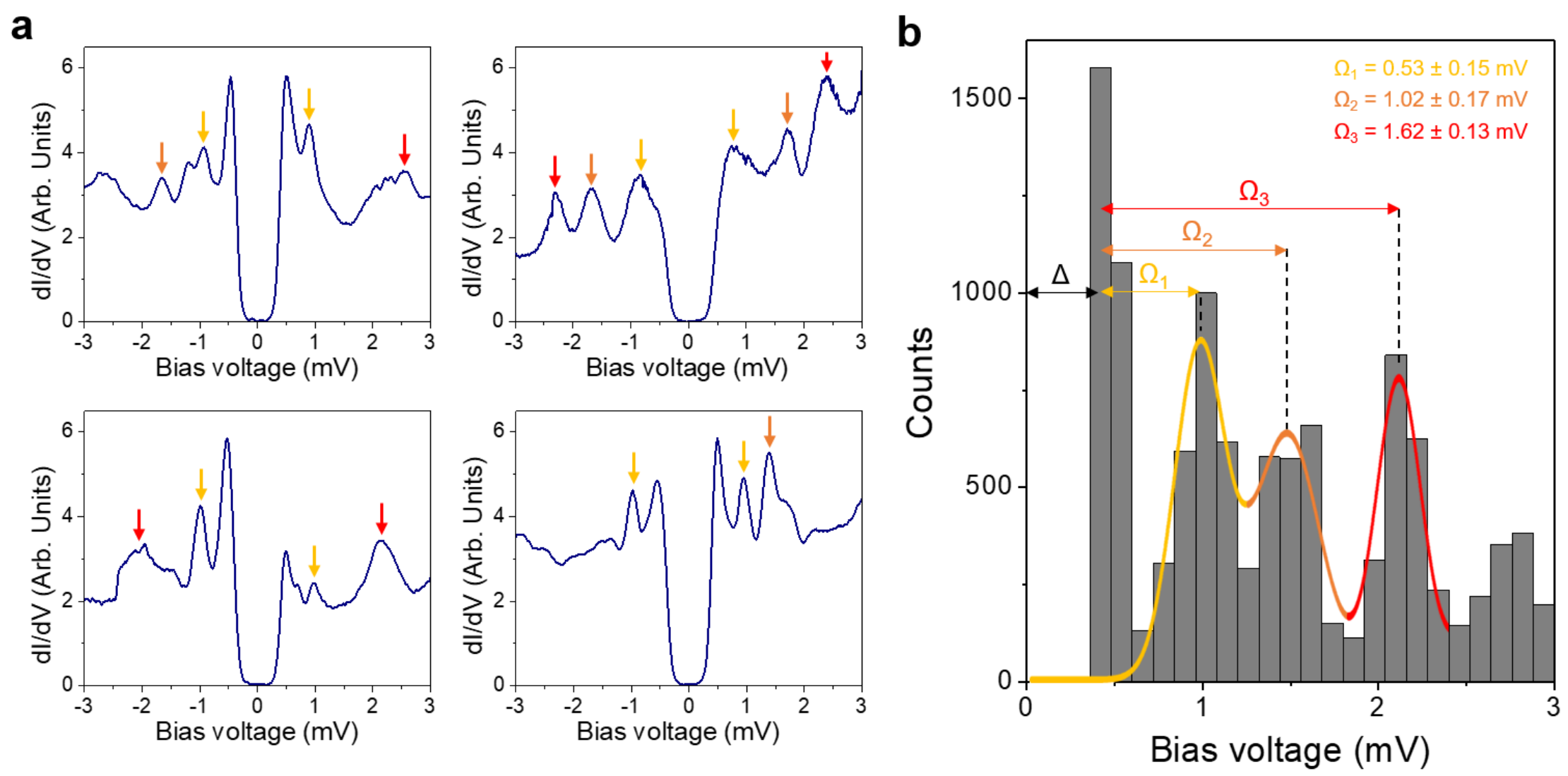

**Figure 2 | Statistical analysis of the STS dip-hump features. a**, Four representative dI/dV curves acquired in single-layer $NbSe_2$ at T = 0.34 K. The arrows identify the fundamental $\Omega_1$ mode (yellow) and the harmonics $\Omega_2$ (orange) and $\Omega_3$ (red). Parameters: f = 833 Hz, $V_{a.c.}$ = 20 µV. **b**, Histogram of 2855 dI/dV curves acquired on different locations, and using different samples and tips. Three clear peaks can be identified for energies larger than the superconducting gap (Δ). A Gaussian fit to the peaks yield the following values: $\Omega_1$ = 0.53 meV, $\Omega_2$ = 1.02 meV and $\Omega_3$ = 1.62 meV.

The non-flat structure of the histogram along with the tip calibration procedures on Cu(111) and graphene (SM) enables to rule out tip effects as the origin for these peaks. The main energy values of the identified peaks ($\Omega_{n=1-3}$), as defined from the nearest coherence peak ($\Omega_n = E_n - \Delta$ with $E_n$ the energy of the n-*th* peak from $E_F$), appear to be in all cases multiple of the energy of the first peak, i.e., $\Omega_n = n \cdot \Omega_1 = n \cdot 0.53\ meV$. Therefore, it appears reasonable to interpret them as harmonics of the same mode $\Omega_1$.

To further characterize the satellite features, we first study their temperature dependence. **Figure 3**a shows a representative dataset of the evolution of the $\Omega_1$ and $\Omega_2$ features as the temperature approaches $T_C \approx 2$ K. As seen, the amplitude of the peaks rapidly decays in all cases, to finally disappear at 1.4 K. Figure. 3b shows that the temperature evolution of the normalized amplitude of the $\Omega_2$ mode (measured from the conductance floor at 2 meV) for empty states (black dots). The amplitude decays faster than what would be expected from thermal broadening (black curve) and, therefore, their disappearance can also be attributed to the weakening of superconductivity itself, suggesting that the satellite peaks are intrinsic to the superconducting state. The disappearance of these features above $T_C$ allows us to rule out other origins for these peaks unrelated to superconductivity such as band structure effects, extrinsic inelastic features and electronic renormalization due to electron-phonon interactions.

Next, we examine the behavior of the satellite peaks under perpendicular magnetic field ($B_\perp$) at 0.34 K. Figure 3c shows a representative dataset of the evolution of the $\Omega_1$, $\Omega_2$ and $\Omega_3$ features in clean regions of $NbSe_2$ for $B_\perp$ up to 2T. Similar to the behavior observed in the T-dependence, these features gradually smear out with the strength of $B_\perp$ as superconductivity weakens and, ultimately, fade out within the mixed state. This further confirms the intrinsic relation between these satellite features and the superconducting state in single-layer $NbSe_2$. We also observe that the maxima of the satellite peaks shift towards higher energies as $B_\perp$ is increased. This is particularly evident for the fundamental mode $\Omega_1$ at both polarities, which shifts in a non-linear fashion as shown in the inset of Figure 3d (See SM for the evolution of the SC gap).

A different way of quantifying the relation between $\Omega_n$ and $\Delta$ is to look at local spatial variations of the superconducting order parameter $\Delta(\vec{r})$, and whether they correlate with the local boson energy $\Omega_n(\vec{r})$, as both are accessible through STS measurements. In **Figure 4**, we show the correlation for the fundamental mode $\Omega_1$ (yellow dots) and higher harmonics $\Omega_2$ and $\Omega_3$ (orange and red dots, respectively) from the set of dI/dV curves used to obtain the histogram of Figure 2b. As seen, all $\Omega_n$ modes exhibit an inverse correlation with respect to $\Delta$ with similar slope (black lines are the linear fits). This observation is consistent with the anticorrelation observed in the study of the $B_{\perp}$-dependence (Figure 3d). A further key insight is the fact that the majority of the values of the fundamental mode $\Omega_1$ are smaller than 2Δ ($\Omega_1/2\Delta < 1$), in contrast to conventional superconductors where phonon-related features frequently lie beyond 2Δ, as in Pb with $\Omega_1/2\Delta \approx 1.7$ (see SM). The statistical confirmation that the fundamental mode has an energy below the pair breaking scale 2Δ implies that this mode cannot decay into fermionic quasiparticles and is therefore undamped, further supporting its interpretation as a superconducting collective mode.

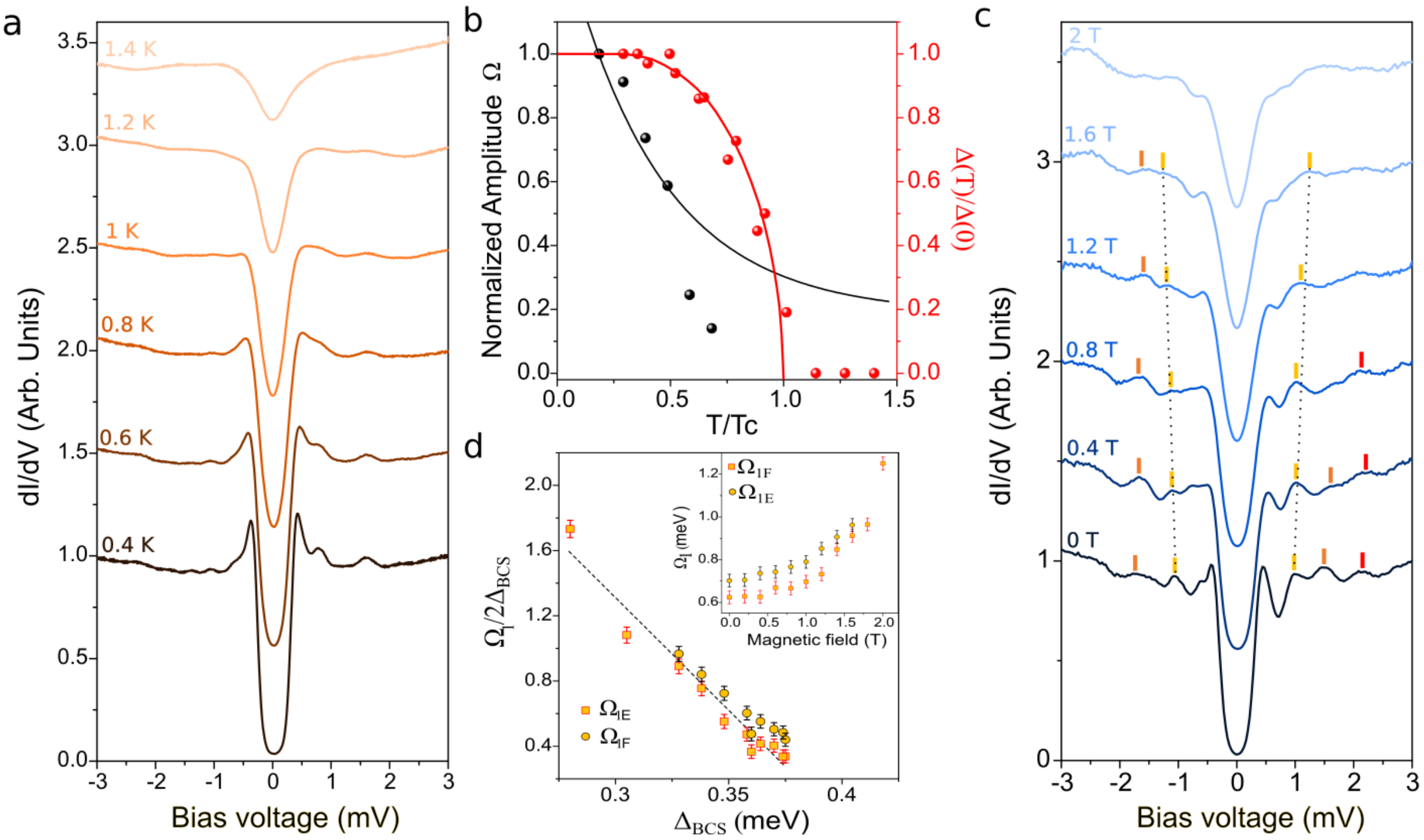

**Figure 3 | Temperature and magnetic field dependence of the bosonic modes. a**, Evolution of the bosonic modes with temperature from 0.4 K up to 1.4 K (f = 833 Hz, $V_{a.c.}$ = 20 μV). **b**, (black dots) Normalized amplitude of the $\Omega_2$ mode for empty states in **a**, showing its decay with T below $T_C \approx 1.9$ K. The decay of amplitude of these STS features from thermal broadening (black curve) is shown to be slower with T. In red, the evolution of the measured Δ with T (circles, fitted to BCS) along with its T-dependence in the BCS theory (line). **c**, Dependence of the bosonic modes with the perpendicular magnetic field ($B_{\perp}$) up to 2 T (f = 833 Hz, $V_{a.c.}$ = 30 μV). Marks indicate the maxima of the resonances and dashed lines connect the energy positions of the fundamental mode $\Omega_1$. **d**, Ratio $\Omega_1/2\Delta$ versus Δ extracted from the $B_{\perp}$-evolution in **c** (Δ here is extracted from the BCS fit). Circles (Squares) represent the filled (empty) states $\Omega_{1F}$ ($\Omega_{1E}$). The dashed line is the linear fit. The inset illustrates the non-linear energy shift of the fundamental mode $\Omega_1$ with the magnetic field.

To summarize our experimental evidence, the STS spectrum of superconducting monolayer $NbSe_2$ displays, in addition to the standard coherence peak at Δ, three satellite peaks at $\Omega_n = \Delta + n\Omega_1$ with $\Omega_1/2\Delta < 1$. These satellites gradually disappear with T and B as the superconducting state weakens, and their position shows a clear statistical anticorrelation between $\Omega_n/2\Delta$ and Δ. These observations are reproduced in two different substrates (graphene and h-BN), which allow us to rule out the potential role of the substrate in the formation of these STS features. Our findings are strong evidence for the presence of a collective mode of energy $\Omega_1$ associated to the superconducting state, which is coupled to fermionic quasiparticles and leaves its imprint in the tunneling spectra (∝ DOS). These observations have important implications regarding the nature of the pairing in this system, which we now discuss.

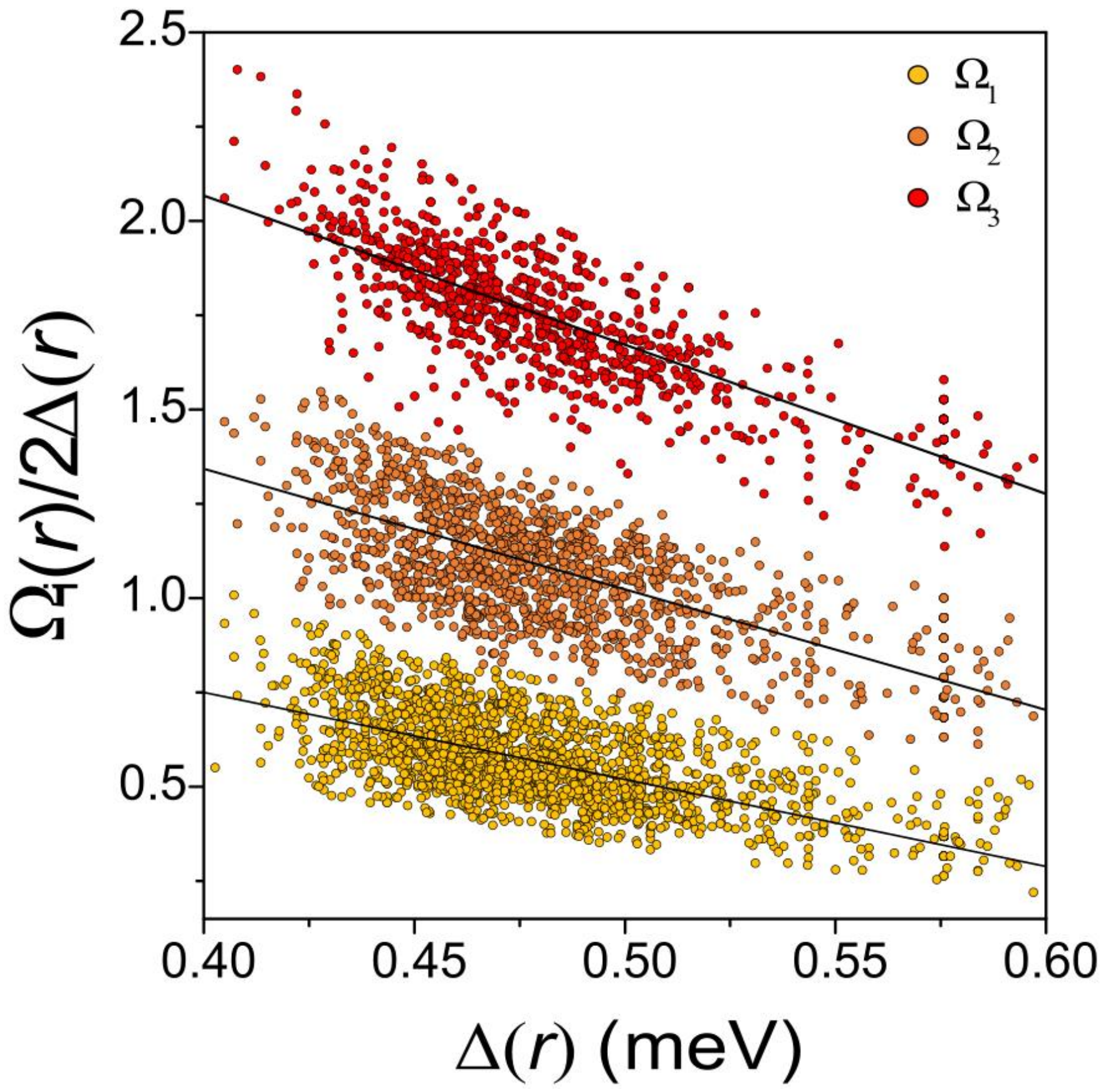

**Figure 4 | Local variations of the bosonic modes and the superconducting gap.** Scatter plot of Ω/2Δ against the superconducting gap (Δ) for the bosonic modes ($\Omega_1$, $\Omega_2$ and $\Omega_3$), showing anticorrelation in all cases. The plot is obtained from the identification of the different Ω modes in 1974 dI/dV curves taken at T = 0.34 K in several samples, and using different calibrated STM tips. The black lines are fits for each subset of points. See SM for details regarding the analysis of the STS data.

## III. Theoretical calculations

The existence of a collective mode can impact the spectral function in two ways, via elastic scattering (the renormalization of the electron self-energy due to virtual boson emission), which leads to a peak at $\Delta+\Omega_1$, as well as inelastic scattering (where quasiparticles might emit real bosons in the tunneling process) which leads to an onset-like feature. While the relative weight of these two contributions is system-dependent, the dip-hump shape of the satellite peaks observed in single-layer $NbSe_2$ closely resembles those features previously observed in strongly correlated superconductors, dominated by elastic scattering, rather than the typical shoulder-dip shape features of conventional superconductors induced by phonons (see the case of Pb(111) in SM) where both contributions can be comparable. Additionally, phonons cannot be responsible for our observed peak since there are none in the relevant energy range. In the high temperature phase without CDW, the phonon spectrum of single-layer $NbSe_2$ shows no relevant features below 3 meV (Ref. [26,39]) and in the presence of the CDW, the lowest CDW phonon mode in the monolayer occurs at 70 $cm^{-1}$ (8.6 meV)[40].

We therefore interpret our peaks as induced by elastic scattering from a collective mode intrinsic to the superconducting state, where two types of collective modes are possible. The first type belongs to excitonic fluctuations (or particle-hole modes), which become sharper after pairing due to the removal of decay channels into fermionic states. These modes might also be the mediators of the interaction that gives rise to superconductivity, or they may be detrimental for it, i.e. pair breaking. A common example in many unconventional superconductors is a resonant magnetic excitation of spin-1[41] (an antiferromagnetic spin-wave) which is believed to mediate superconductivity[42] in cuprates[3–10], Fe-pnictides[11–16] and heavy fermion compounds[20,21]. Another known example are nematic fluctuations, as found in the Fe superconductors[43]. The second type are superconducting fluctuations (or particle-particle modes), most commonly due to close competition between pairing channels, like Leggett modes[44] in two-band superconductor $MgB_2$ or Bardarsis-Shrieffer (BS) modes[45] in Fe superconductors where s-wave and d-wave pairings are close competitors[46]. Either type of collective mode can be observed with different experimental techniques[47–49], including tunneling experiments, where bosonic modes are identified via the mentioned characteristic dip-hump features[8,10,12,14,15,17–20]. While these STS experiments are mostly interpreted in terms of particle-hole modes like spin-waves, there is no reason to preclude particle-particle modes to be found with this technique. Finally, all superconductors have an amplitude or Higgs mode, which is normally unobservable on its own[50], but it has been observed in bulk $NbSe_2$ due to its mixing with collective CDW modes[51]. Nevertheless, the Higgs mode can readily be discarded because in monolayer $NbSe_2$ the CDW mode has much higher energy than $2\Delta$ and their coupling is highly suppressed.

Which of the previous collective mode scenarios applies to our experiment? Monolayer $NbSe_2$ has been predicted to be near a ferromagnetic instability[25,26] which competes with the CDW and, therefore, spin fluctuations could be sizable and potentially give rise to a particle-hole collective spin-wave. Such mode would indeed broaden and disappear as the temperature or magnetic field are increased to their critical values as observed in cuprates[47] and Fe-based materials[48]. Nevertheless, no magnetic order has been found in $NbSe_2$, and there is no direct evidence of strong spin fluctuations either. In the particle-particle scenario, however, there is a very natural mechanism for the emergence of collective modes: the competition between pairing channels signaled by the emergence of magnetic field-induced nematic superconductivity. To substantiate the characteristics of these collective modes, we now present a microscopic model of this competition which leads to explicit predictions that can be compared with our experiment.

$NbSe_2$ bands near the Fermi level are derived from the three $t_{2g}$ Nb *d* orbitals, and consist of a hole pocket around the Γ point with dominant $d_{z2}$ character and hole pockets around the K points with $d_{x2-y2}\pm i d_{xy}$ character. This difference leads to strong Ising SOC for the K pockets but negligible SOC for the Γ pocket, and to different k-independent pairing channels: while both Γ and K pockets admit the standard *s*-wave state, the K points can also develop spin-triplet, orbital-singlet pairing of the $d_{x2-y2}, d_{xy}$ orbitals which has *f*-wave symmetry[52]. For simplicity, we therefore assume the Γ pocket is a spectator with *s*-wave symmetry gap and use a model with just the K pockets

$$H_0(k) = \Psi^\dagger \left[ (-\frac{k^2}{2m} - \mu)\tau_0\sigma_0 + \lambda\tau_z\sigma_z \right] \Psi$$

where the $\tau_i$ and $\sigma_i$ matrices act on the valley and spin index respectively, and $\lambda$ is the Ising SOC. The pairing operators can be written as $\Delta_S = \Psi\tau_x i\sigma_y\Psi$ for the *s*-wave singlet which has $A'_1$ symmetry, and $\Delta_T^i = \Psi\tau_y\sigma_y\sigma_i\Psi$ with $i = x, y, z$ for the *f*-wave triplet, where $\Delta_T^z$ belongs to an $A'_1$ irrep while $\Delta_T^{x,y}$ make an $E''$ irrep (see **Figure 5**a for a schematic). In the presence of SOC, the mixing of the $A'_1$ singlet and $A'_1$ triplet becomes allowed. This mixing scales with the difference of the DOS of the spin-split pockets which is however very small. In our model with the leading k-independent SOC $\lambda$ the DOS difference and the mixing actually vanish, and the only effect of $\lambda$ is to disfavor the $E''$ state. Nevertheless, if attraction in the *f*-wave channel is sizable, its $E''$ part can naturally be induced with an in-plane magnetic field, which can explain the previous experiments proposing the competition of nematic[36,37] and triplet[38] pairing.

Assuming an $s$-wave ground state and vanishing singlet-triplet mixing, the imaginary fluctuations towards the two nearby $f$-wave triplets $A'_1$ and $E''$ represent two collective modes of the Bardarsis-Schriefer type. The fluctuation towards the $E''$ channel is likely unobservable in practice because $\lambda \gg \Delta$, which implies $T_c$ for the $E''$ state will nearly vanish. We therefore consider only the fluctuation towards the $A'_1$ triplet. In the presence of singlet-triplet mixing, this second mode still exists but no longer has a well-defined Bardarsis-Schriefer character, because the gaps in the spin-split Fermi surfaces take the mixed form $\Delta_{\pm} = \Delta_S \pm \Delta_T^z$. This mode can alternatively be interpreted as the relative phase fluctuation of the $\Delta_{\pm}$ gaps, i.e. a Leggett mode[53] (see Figure. 5b), which we take as the leading candidate to explain our experiments.

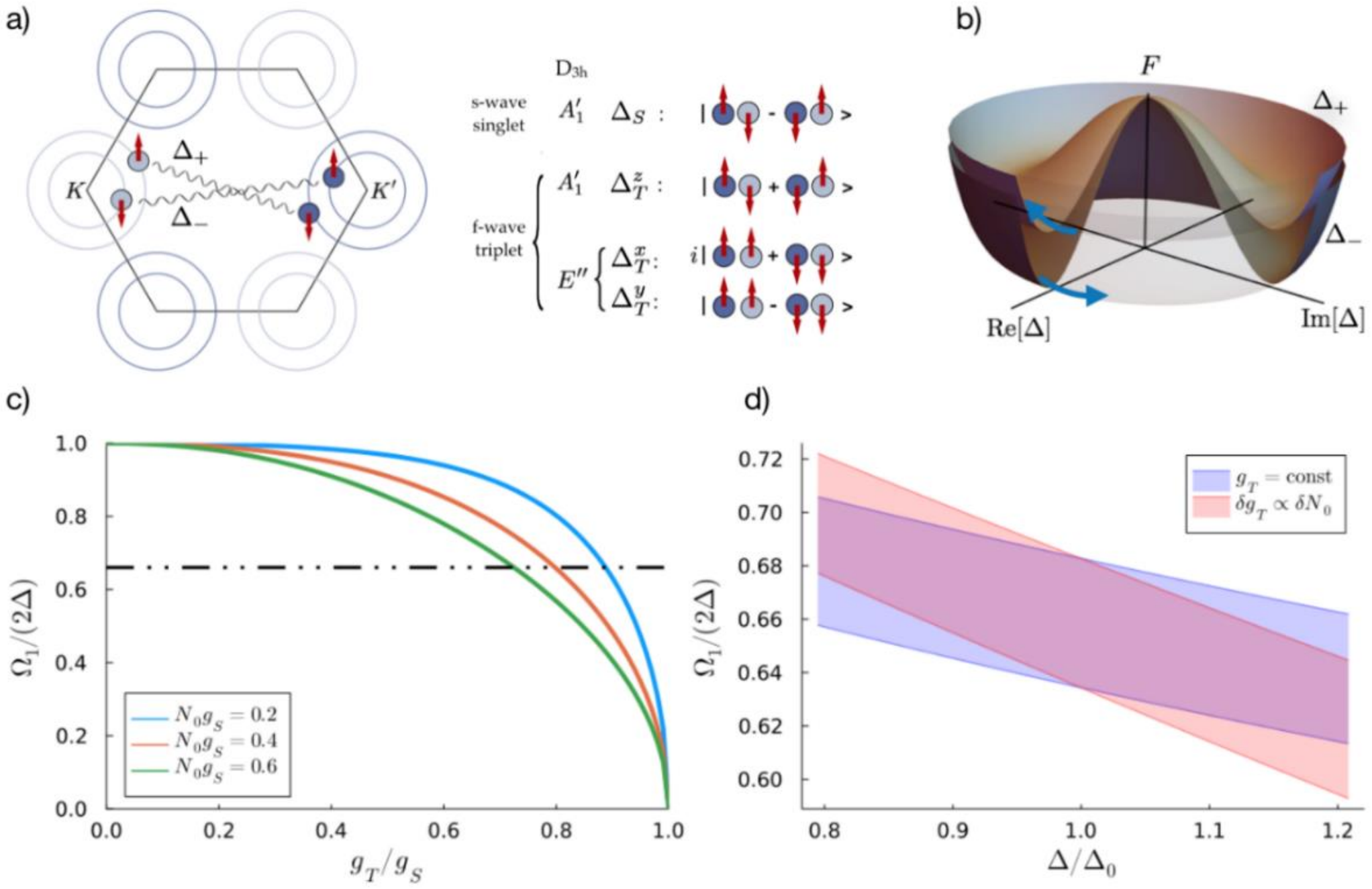

**Figure 5 | Pairing states and collective modes. a**, Schematic Fermi surface near the K points and structure of the different pairing states considered. **b**, Schematic free energy for the gaps of the two spin-split Fermi surfaces, illustrating the Leggett mode as their relative phase fluctuation. **c**, Leggett mode energy normalized by the zero temperature gap $\Omega_1/2\Delta$ as a function of the ratio $g_S/g_T$. The collective mode is gapless when the two couplings are the same, and approaches $2\Delta$ for vanishing triplet attraction $g_T = 0$. **d**, Allowed values of $\Omega_1/2\Delta$ when $N_0$ takes a range of values keeping $g_S$ and $g_T$ constant (blue), and when we additionally assume that $g_T$ is correlated with $N_0$. Anticorrelation is generically observed but is more pronounced in the latter case.

To model the Leggett mode explicitly, we consider attractive interactions in the s-wave singlet and f-wave triplet channels as follows

$$V = g_S \Delta_S^{+}\Delta_S + g_T\Delta_T^{+}\Delta_T$$

with $g_S, g_T < 0$. As discussed in Ref.[52], $g_S$ might be thought of as induced by electron-phonon coupling, while $g_T$ can have contributions from both electron-phonon and Coulomb interactions, in particular through effective spin fluctuation[25]. Depending on whether $|g_S|$ or $|g_T|$ is largest, we have a ground state with dominant singlet or triplet character, and we assume $|g_S| > |g_T|$. The energy of the Leggett mode can be computed with this model (see SM) and is obtained from the implicit equation

$$\frac{arcsin\,\Omega_1/2\Delta}{\sqrt{(2\Delta/\,\Omega_1)^2-1}} = \frac{1}{N_0|g_T|} - \frac{1}{N_0|g_S|}$$

From the measured value of $\Omega_1/2\Delta = 0.66$ we can estimate the ratio $g_T/g_S$ and hence how close the triplet state is. This first requires an estimate of $N_0 g_S$. If we assume a weak coupling BCS limit, a gap of $\Delta \sim 0.4$ meV and a Debye frequency cutoff in the range of bulk estimates $\omega_D \sim 20$ meV (ref. [54]), this corresponds to $N_0 g_S \sim 0.2$. However, since the ratio $2\Delta/k_B T_c \sim 4.9$ in our experiment would denote moderate to strong coupling, the value of $N_0 g_S$ is likely larger. Figure 5c displays the numerical solution of $\Omega_1/2\Delta$ as a function of $g_T/g_S$ for $N_0 g_S = 0.2 - 0.6$, showing this produces a range $g_T/g_S$ = 0.7-0.9. The triplet attraction must therefore be sizable, but still not enough to overcome the singlet attraction.

To show that this collective mode can in fact be observed in STS measurements, we have also computed the tunneling spectra due to the renormalization of the fermionic self-energy by this collective mode following Ref.[55], showing that it indeed leads to a peak at $E_1 = \Omega_1 + \Delta$ (See SM). This calculation could be extended to higher orders to show the existence of harmonics at $E_n = \Omega_1 + n\Delta$ as well. A prediction of the absolute amplitude of the peaks is however beyond the scope of our calculation.

Our theory also allows to predict that the energy of the collective mode has a similar exponential dependence on temperature as the gap itself (see SM). Because of this, the collective mode energy should stay roughly constant in T for low T as we observe and only show deviations as it approaches $T \sim T_c$, where estimating the energy is prevented by our resolution. Similarly, the amplitude of the peak is rapidly suppressed near $T \sim T_c$ because the weight of this boson, computed as the residue of its propagator, scales $\propto \Delta^2$. In the presence of a magnetic field, the collective mode energy shows a significant rise, surpassing $2\Delta$ even at moderate fields B~1T. While a quantitative prediction for this would involve modeling the vortex mixed state, it is clear that this change cannot originate just from changes in the gap, and we conjecture that the magnetic field might reduce $g_T$ by hardening spin fluctuations. Complementary probes of this collective mode are needed to better understand its behavior under magnetic fields.

To address the observed local anticorrelation with the gap, we assume that local variations of the model parameters lead to variations in the collective mode energy[56]. Figureure 5d shows the predicted band of allowed energies for two different scenarios. First, we consider that $N_0$ varies spatially, leading to variations of $\Delta$, while $g_S$ and $g_T$ are kept constant. Moderate anticorrelation is obtained in this case. If we further assume that $g_T$ depends on the DOS, as it would be for example if it relied on spin-fluctuations, we see that a larger anticorrelation is attained. Analysis of other scenarios shows the anticorrelation is quite generic for this collective mode, while a detailed match with experiments will require exact knowledge of the origin of the spatial fluctuations. Overall, we believe our model supports our hypothesis that the observed mode is the Leggett mode due to proximity of *f*-wave triplet and provides a consistent picture for our observations.

**IV. Discussion**

Finally, it is also interesting to compare the case of single-layer $NbSe_2$ with that of other superconductors where particle-hole magnetic resonances have been observed, where there is an empirical universal relation between resonance energy and the gap as $\Omega/2\Delta \sim 0.64$ over two orders of magnitude of $\Delta$ (ref.[57]). In this context, single-layer $NbSe_2$ lies in the region of the smallest $\Omega$'s along with the heavy-fermion compounds[20,21] with a very similar value $\Omega_1/(2\overline{\Delta_{BCS}}) = \frac{0.53}{0.8} = 0.66$. Such intriguing similarity invokes further comparative investigation between particle-hole and particle-particle collective modes.

In summary, our results in single-layer $NbSe_2$ have unequivocally demonstrated the existence of a bosonic, undamped collective mode associated to the superconducting state, which we have interpreted as the fluctuations to a competing *f*-wave triplet channel. Our findings create exciting new opportunities for directly exploring unconventional superconductivity in a 2D material of simple synthesis, handling, and experimental analysis. We expect that this work will trigger active research in other simple 2D TMD superconductors, where competing superconducting channels and eventually triplet superconductivity could arise as well.

**Acknowledgements**

We acknowledge fruitful discussions with Félix Ynduráin. We also thank Samuel Navas for providing us with bulk $NbSe_2$ crystals. M.M.U. acknowledges support by the ERC Starting grant LINKSPM (Grant 758558) and the Spanish MINECO under grants no. PID2020-116619GB-C21. D.M.S. is supported by a FPU predoctoral contract from MEFP No. FPU19/03195. F. J acknowledges funding from the Spanish MCI/AEI/FEDER through grant PGC2018- 101988-B-C21 and form the Basque government through PIBA grant 2019-81.

# Supplementary Materials for

# Observation of superconducting collective modes from competing pairing instabilities in single-layer $NbSe_2$

*Wen Wan*[1,†] *Paul Dreher*[1,†,] *Daniel Muñoz-Segovia*[1], *Rishav Harsh*[1], *Haojie Guo*[2], *Antonio J. Martínez-Galera*[2,3], *Francisco Guinea*[1,4], *Fernando de Juan*[1,5] *and Miguel M. Ugeda*[1,5,6,*]

CONTENTS

---

* These authors contributed equally to this work.

† Corresponding author: mmugeda@dipc.org

## I. GROWTH OF SINGLE-LAYER $NbSe_2$

Single-layer $NbSe_2$ was grown on epitaxial BLG on 6H-SiC(0001) by molecular beam epitaxy (MBE) at a base pressure of ∼5×10-10 mbar in our home-made UHV-MBE system. SiC wafers with resistivities $\rho \sim 120\ \Omega$ cm were first cleaned with an isopropyl solution in an ultrasonic bath. Thereafter, they were put into a UHV-MBE chamber and annealed at a temperature of 700 °C for 1h for outgassing. Then, the graphitization of the SiC surface was carried out using an automatized cycling mechanism where the sample was ramped between 700°C and 1350°C at a continuous ramping speed of ∼20°C/s. The SiC crystal was kept for 30s at 1350°C. In total, 80 cycles were performed[1]. Reflective high energy diffraction (RHEED) was used to monitor the layer growth progression from the SiC to the final $NbSe_2$ layer (S. Figs. 1a,b). During the growth of $NbSe_2$, the obtained BLG substrate was kept at 570°C. High purity Nb (99.99%) and Se (99.999%) were evaporated using an electron beam evaporator and a standard Knudsen cell, respectively. The Nb:Se flux ratio was kept at 1:30, while evaporating the Se led to a pressure of ∼5 × 10-9 mbar inside the UHV chamber (Se atmosphere). Samples were prepared using an evaporation time of 30 min in order to obtain a coverage of ∼0.8 ML. Subsequently, evaporation of Se was kept for additional 5 min in order to minimize the presence of atomic vacancies. Atomic Force Microscopy at ambient conditions was routinely used to optimize the morphology – island and domain sizes, coverage and cleanliness of the $NbSe_2$ film (AFM image of S. Fig. 1c). The samples used for AFM characterization were not further used for STM. In order to transfer the samples from our MBE to the STM system, they were capped with a ∼10 nm film of Se to protect them against oxidation. The capping layer was easily removed in the UHV chamber of the STM by annealing the sample at ∼250°C.

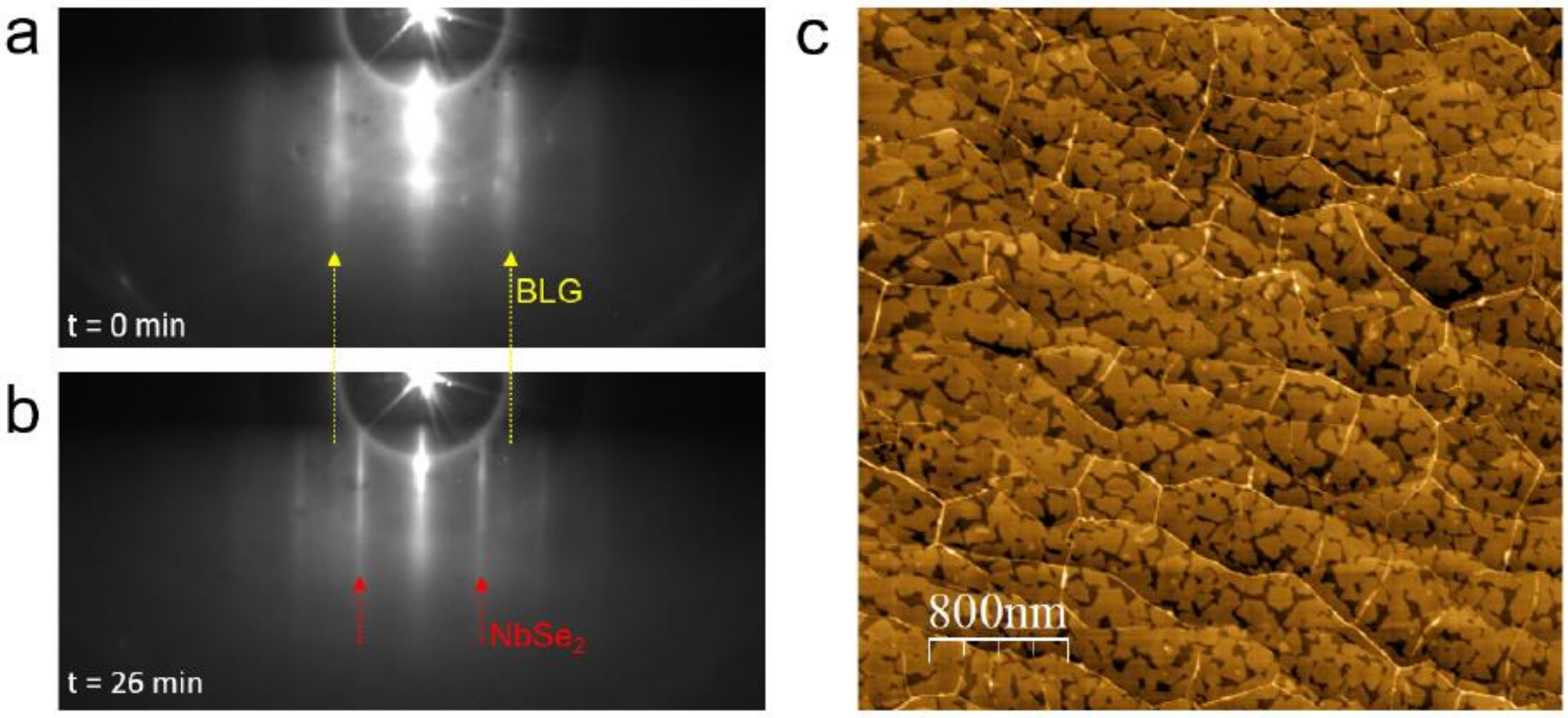

**Supplementary Figure 1: Single-layer $NbSe_2$ growth.** a) RHEED pattern after the growth of BLG on SiC(0001). Yellow lines indicate the diffraction features of BLG. In the course of the evaporation $NbSe_2$ is formed on the surface, thus gradually appearing new diffraction lines (red arrows) and attenuating those of BLG, as shown in b). c) Amplitude-modulation AFM image of a typical single-layer $NbSe_2$ on BLG/SiC(0001) sample used here.

## II. STM/STS MEASUREMENTS AND TIP CALIBRATION

Scanning tunneling microscopy and spectroscopy on single-layer $NbSe_2$, bulk $NbSe_2$ and Pb(111) experiments were carried out in an ultra-high vacuum (UHV), low temperature and high magnetic field scanning tunneling microscope USM-1300 (Unisoku Co., Ltd.) operated at T = 0.34 K. STS measurements were performed using the lock-in technique with typical a.c. modulations of 20-50 $\mu$V at 833 Hz. We used Pt/Ir tips for the STM/STS experiments. STM/STS data were analyzed and rendered using WSxM software.[2]

To avoid tip artifacts in our STS measurements, the STM tips were systematically calibrated using a Cu(111) surface as reference. S. Fig. 2a shows a typical dI/dV curve after tip calibration showing the onset of the surface state of Cu(111) at of -0.44 eV. We also performed careful inspection of the DOS around EF to avoid the use of functionalized tips showing strong variations in the DOS. Furthermore, simultaneous comparative dI/dV curves were regularly done on the graphene substrate in between STS experiments in NbSe2. Fig. 2b shows an example of these consecutive control measurements on graphene, where the peaks are clearly absent and the only reproducible feature is a proximity-induced superconducting gap. We point out that these procedures along with the non-flat structure of the histogram (Fig. 2b in the manuscript) enable to safely rule out tip excitations as the origin of the observed resonances.

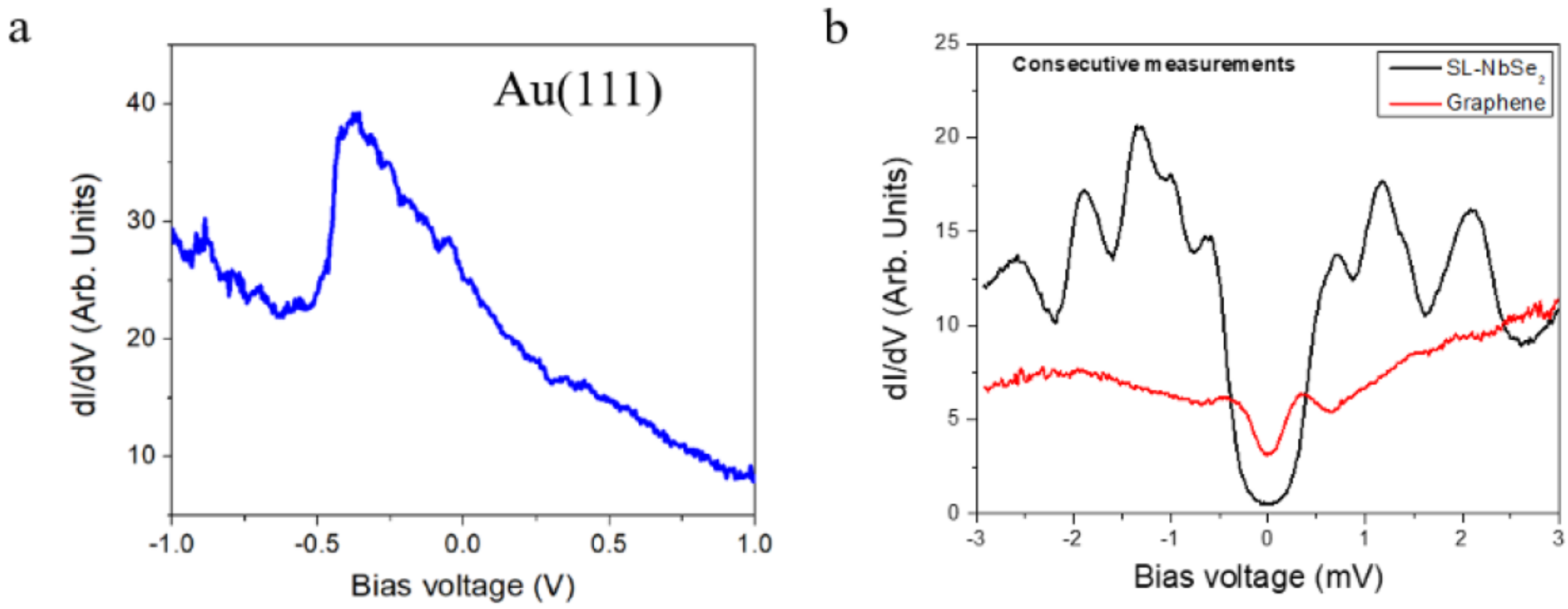

**Supplementary Figure 2:** Tip calibration. a) Typical dI/dV spectrum taken during the calibration of a PtIr on Cu(111) (f = 833 Hz, Va.c. = 2 mV, T = 4.2 K). b) Consecutive dI/dV curves measured in NbSe2/BLG (black) and BLG (red).

## III. BOSONIC MODES IN SINGLE-LAYER NBSE$_2$ ON H-BN: THE ROLE OF THE SUBSTRATE

In order to evaluate the potential role of the substrate in the emergence of the observed peak features, we have studied the quasiparticle spectrum near the SC gap of SL-NbSe$_2$ grown on h-BN on Ir(111) as it was done on graphene (main manuscript). Single-layer h-BN has a 6 eV gap, which enables the study of electronically decoupled low dimensional systems[3,4]. Single-layer NbSe2 shows a superconducting gap of 0.39 meV on h-BN2, a nearly identical with the value obtained on NbSe$_2$ on graphene, which is a first indication of the negligible role of the substrates in the superconductivity of NbSe$_2$. Supplementary Figure 3a shows a representative dI/dV spectrum taken on SL-NbSe$_2$ on h-BN at 0.34 K. As can be seen, the spectrum shows similar resonances (black arrows) near the coherence peaks as in NbSe$_2$ on graphene case. We have also done statistical analysis of the energy position of these peaks. S. Fig. 3b shows the resulting histogram of peaks obtained from a set of 118 dI/dV spectra taken in SL-NbSe$_2$/h-BN in different regions and using different tips. The histogram also shows three clear equidistant peaks, which fitted by Gaussian distributions yield $\Omega_n = n \cdot \Omega_1 = 0.41n \pm 0.14$ meV, with $n = 1, 2, 3$, in good agreement with what is observed in SL-NbSe$_2$/graphene (Fig.2b in the manuscript). Furthermore, this collection of peaks show the expected anticorrelation as shown in S. Fig 3c. Lastly, we have measured the evolution of these peaks with the magnetic field, as shown in S. Fig. 3d. Here the peaks are also gradually weakened with B and disappear within the mixed state. In conclusion, we have reproduced the same phenomenology of SL-NbSe$_2$ on two electronically distinct substrates, which unarguably demonstrates their negligible role in the formation of the peaks in the quasiparticle spectrum in this 2D superconductor.

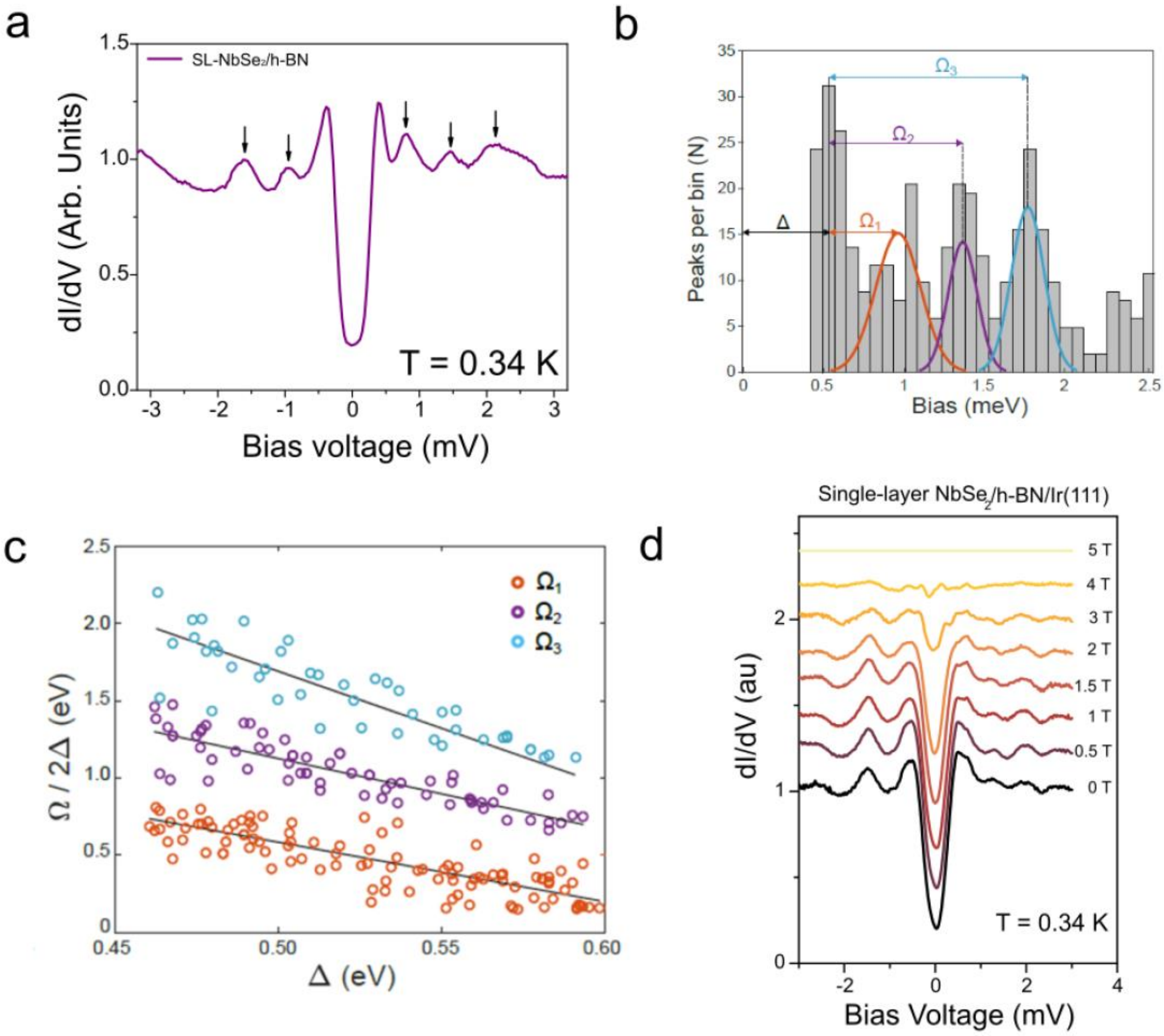

**Supplementary Figure 3:** Single-layer NbSe$_2$ on h-BN/Ir(111). a) Typical dI/dV spectrum taken on single-layer NbSe$_2$/h-BN (f = 833 Hz, $V_{a.c.}$ = 40 $\mu$V, T = 0.34 K). The black arrows show the peaks beyond the SC gap. b) Histogram of 118 dI/dV curves acquired on different locations, and using different samples and tips. Three peaks can be identified beyond the superconducting gap ($\Delta$). A Gaussian fit to the peaks yield the following values: $\Omega_1 = 0.41 \pm 0.14$, $\Omega_2 = 1.02 \pm 0.14$, $\Omega_3 = 1.62 \pm 0.14$. c) Scatter plot of $\Omega/2\Delta$ against $\Delta$ for the bosonic modes ($\Omega_1$, $\Omega_2$, $\Omega_3$), showing anticorrelation in all cases. d) Evolution of the peaks in single-layer NbSe$_2$/h-BN with the magnetic field at T = 0.34 K.

## IV. ENERGY DETERMINATION OF THE BOSONIC MODES

The energy positions of bosonic modes in previous STS measurements are usually taken either as the maxima/minima of the $d^2I/dV^2$ spectrum or the maxima of the dI/dV spectrum. Although the $d^2I/dV^2$ analysis is more used due to the involved inelastic tunneling component, we have chosen the latter option on behalf of simplicity in the statistical analysis. Numerical derivative of the dI/dV spectra could introduce a non-negligible error in the energy determination of the peaks, thus likely making both methods equally accurate. The $d^2I/dV^2$ method is particularly suited for systems where the dI/dV signal is dominated by a strong background and no maxima can be identified, which is not our case. Nevertheless, the sharpness of the bosonic modes in this system leads to systematic energy differences between both methods are in the range 50-100 $\mu$eV for all $\Omega$n modes and $\Delta$ (see S. Fig. 4). Therefore, the choice of the method to determine the energy position of the bosonic modes does not affect the $\Omega$ vs. $\Delta$ anticorrelation and $\Omega/2\Delta$ ratios.

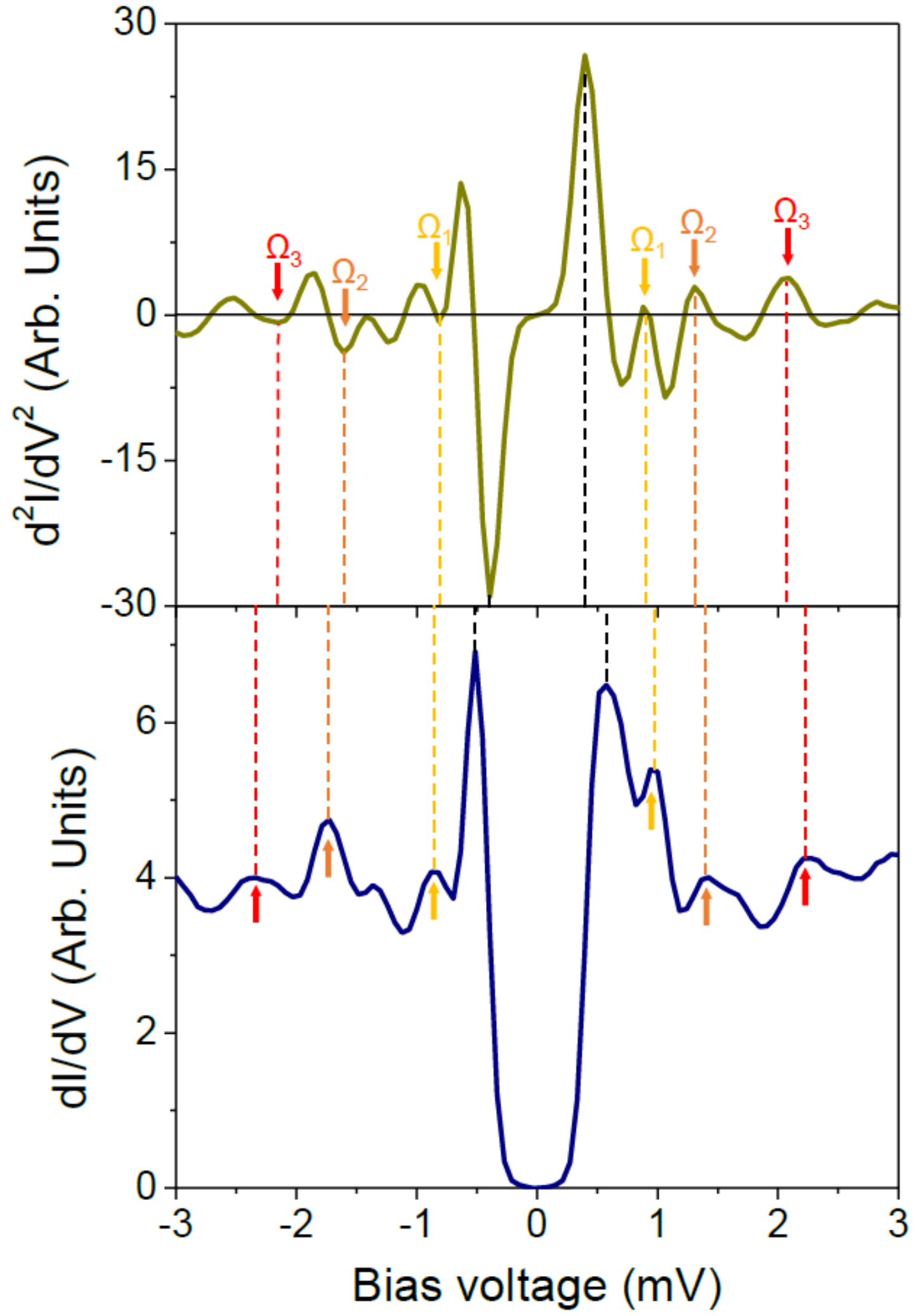

**Supplementary Figure 4:** The dI/dV versus $d^2I/dV^2$ determination of the bosonic modes. Comparison between the energy determination of the STS features in single-layer $NbSe_2$ in the dI/dV spectrum (Lower panel) and in the $d^2I/dV^2$ of the same curve (Upper panel). (f = 833 Hz, $V_{a.c.}$ = 20 $\mu$V).

## V. STATISTICAL ANALYSIS OF THE BOSON MODE ENERGIES

We performed statistical analysis of the STS data to obtain the histogram and the $\Omega$ vs. $\Delta$ plot shown respectively in Fig. 2b and Fig. 4 of the main text. We used an ensemble of 2855 dI/dV spectra (T = 340 mK) accumulated over tens of different regions from five different samples and using multiple different calibrated tips. For the identification of the peak satellites in the dI/dV curves, we used a MATLAB script that includes the built-in function findpeaks. We defined the peaks as specific height differences in neighboring data point minima, considering also the width of their rising/falling slope. By normalizing the set-point conductance of all STS curves to 1, a threshold height of $h_{min} = 0.03$ and peak width of $w_{min} = 0.1$ meV were used to remove noise and, therefore, false peaks. We observe no significant changes in the histogram while varying the bin size and the spin resonances $\Omega$n are always clearly prominent.

To obtain the correlation between the SC gap size and the spin resonance energy (plot in Fig. 4 of the main text), we used a similar MATLAB script to identify the modes in each dI/dV curve. We chose a limit of data point restriction within $2\Theta$ for each $\Delta$ (twice the standard deviation) as threshold values below $\Omega_1$ and above $\Omega_3$. dI/dV curves with multiple peaks within $4\Theta$ around each $\Omega_n$ value were not used in the correlation analysis. (1974 curves were used in total). The Pearson correlation coefficient for the spin resonances to yield values of $P\Omega_1$ = - 0.45, $P\Omega_2$= - 0.35, and $P\Omega_3$ = - 0.19.

## VI. TUNNELING SPECTROSCOPY IN BULK $NbSe_2$

Tunneling spectroscopy in bulk $NbSe_2$ at milikevin temperatures were initially performed by Hess, et al.[5,6], and later by Guillamón, et al.[7]. These measurements did not show any signature of boson modes in bulk $NbSe_2$ at temperatures as low as 50 mK. We have corroborated this fact by measuring the superconducting gap in bulk $NbSe_2$ using calibrated tips (see S. Fig. 5).

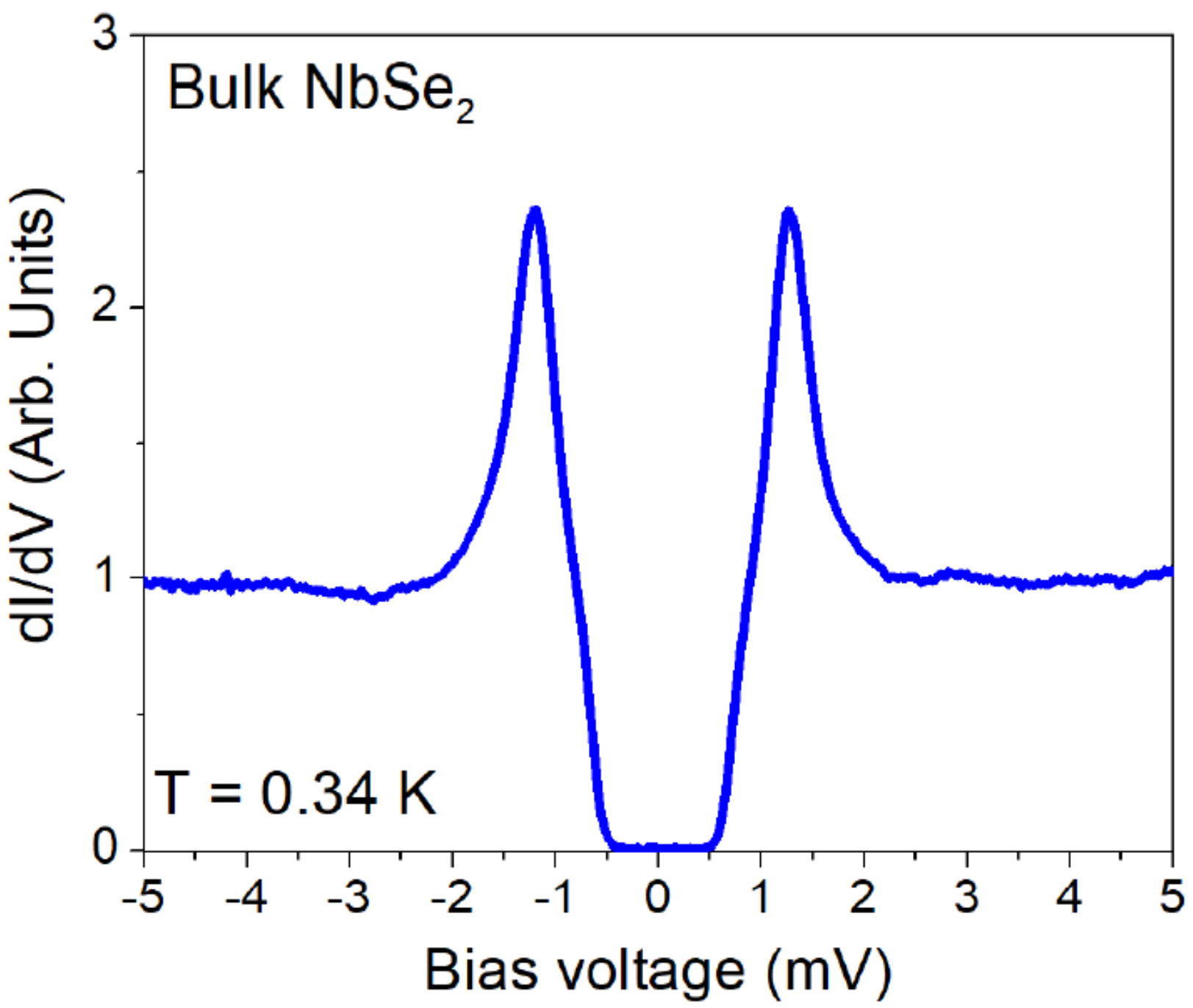

**Supplementary Figure 5:** STS in bulk $NbSe_2$. Typical dI/dV curve in bulk $NbSe_2$ acquired at T = 0.34 K (f = 833 Hz, $V_{a.c.}$ = 30 $\mu$V).

## VII. PHONON FEATURES IN CONVENTIONAL SUPERCONDUCTORS: Pb(111)

Conventional superconductors show signatures of quasiparticle coupling to bosonic modes of phononic origin. These signatures appear in STS measurements as shoulder-dip features rather than the distinctive dip-hump features of unconventional superconductors with Cooper pairing mediated by spin fluctuations. S. Fig. 6 shows these features for the case of Pb(111). Note that the phonon features in this case lie beyond $2\Delta$ in contrast to the dip-hump features in unconventional superconductors.

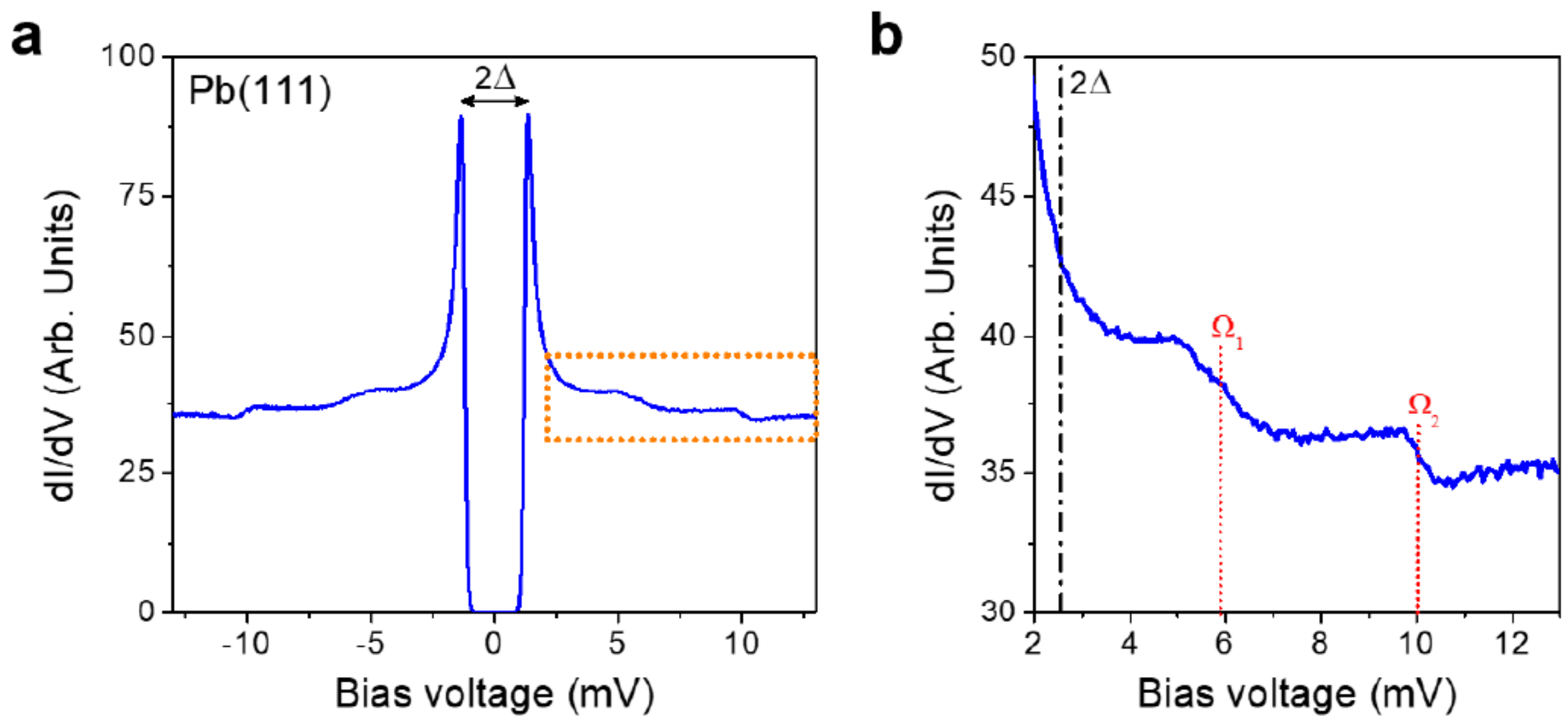

**Supplementary Figure 6: Virtual phonon features in conventional superconductors.** a) STS spectrum acquired around the superconducting gap in Pb(111) at T = 0.34 K (f = 833 Hz, $V_{a.c.}$ = 50 $\mu$V). b) Close-up view of the boxed region in a where the bosonic modes are seen (shoulder-dip features labelled as $\Omega_1$ and $\Omega_2$). Note that both features lie well beyond $2\Delta$.

## VIII. FURTHER DATA ON MAGNETIC FIELD DEPENDENCE

Supplementary Fig. 7 shows the magnetic field dependence of the fundamental mode $\Omega_1$ and the SC gap $\Delta_1$ for the data-set shown in Figure 3c in the main manuscript. $E_1$ is the energy of the mode measured with respect to the Fermi level and, therefore, $E_1 = \Omega_1 + \Delta_1$.

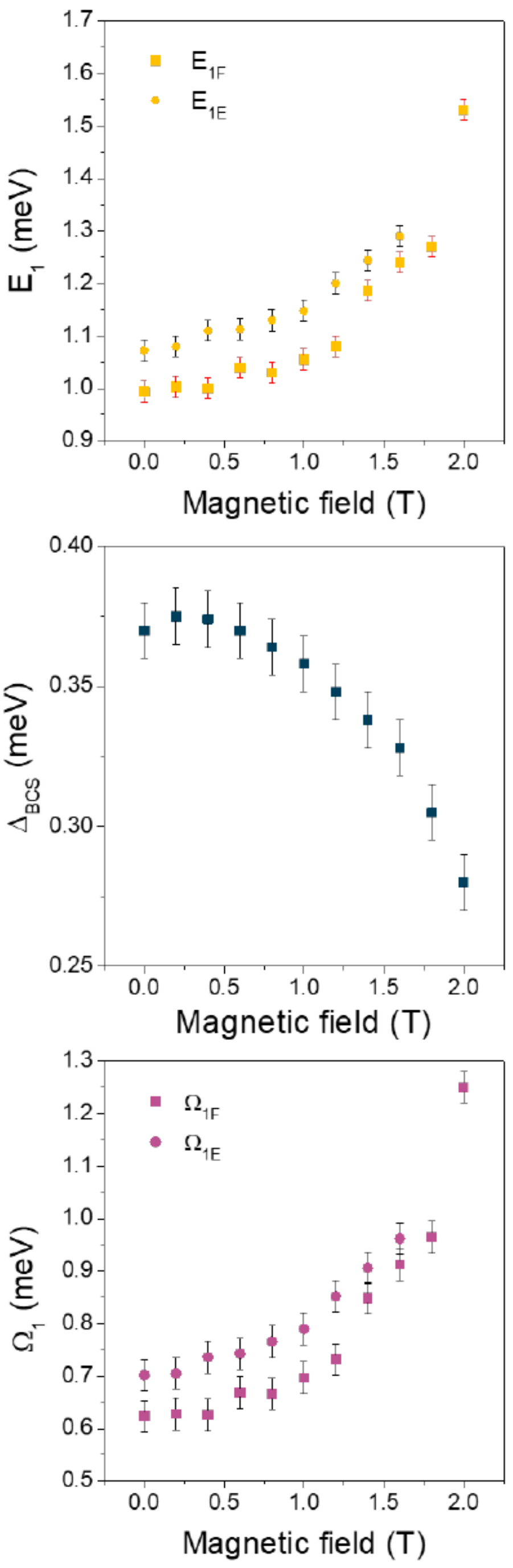

**Supplementary Figure 7: Magnetic field dependence of the main collective mode.**

## IX. THEORY

### A. Model for superconductivity in $NbSe_2$

In this section we summarize the main aspects of the electronic structure of monolayer $NbSe_2$, propose an effective model for superconductivity and discuss the different interactions and pairing channels. Monolayer $NbSe_2$ has a crystal structure with $D_{3h}$ point group, which is generated by $z$-axis threefold rotations $C_{3z}$, $x$-axis twofold rotations $C_{2x}$ and a horizontal mirror plane $\sigma_h$. This group has one-dimensional irreps $A_1'$, $A_2'$, $A_1''$ and $A_2''$, and two dimensional irreps $E'$ and $E''$. The bands near the Fermi level are made of large hole pockets around the $\Gamma$, $K$ and $K'$ points, which derive from the three Nb d orbitals $d_0 = d_{z^2}$ with symmetry $A_1'$ and $(d_1, d_2) = (d_{x^2-y^2}, d_{xy})$ with symmetry $E'$. The $\Gamma$ pocket has dominant $d_0$ character, while the $K, K'$ pockets have dominant $d_\pm = d_1 \pm i d_2$ character. Because of this different character, spin-orbit coupling leads to a large Ising splitting at the $K$ and $K'$ points, while the splitting around the $\Gamma$ point is negligible.

The different pairing channels in this multiband superconductor have been discussed before[8,9]. It is instructive to start the discussion without spin-orbit coupling, and for simplicity to restrict to momentum-independent pairing within each band. In this case, the $\Gamma$ pocket can only have singlet pairing with s-wave orbital symmetry ($A_1'$). However, the $K$ and $K'$ points can have both singlet pairing with s-wave symmetry ($A_1'$, same gap in the two pockets) and triplet pairing with f-wave orbital symmetry (f-wave corresponds to an $A_2'$, opposite gap in the two pockets). When spin-orbit coupling is included, channels cannot be classified independently by orbital and spin symmetry. Given that spin bilinears transform as $S_z \sim A_2'$ and $(S_x, S_y) \sim E''$, the f-wave triplet with out of plane $d$-vector transforms as $A_1'$ while the triplet with in-plane $\vec{d}$-vector transforms as $E''$. The out of plane triplet has the same symmetry as the standard s-wave singlet and can therefore mix with it.

In this work, we have assumed that the superconducting ground state is the standard s-wave singlet, while the f-wave triplet is a close subleading competitor. In this picture, superconductivity in the $\Gamma$ pocket is a spectator, in the sense that it just assumes a fully gapped s-wave state with the same sign as the $K$ pockets due to a residual interband coupling. Because of this, to simplify the discussion we only consider a model for the $K$ and $K'$ pockets, where all the previous statements can be illustrated explicitly. Taking fermionic operators in the band basis as $\psi = (d_{K\sigma}, d_{K'\sigma})$ for the two pockets, we consider the non-interacting Hamiltonian

$$H_0(\boldsymbol{k}) = \psi^\dagger \left[\xi(\boldsymbol{k})\tau_0\sigma_0 + \lambda\tau_z\sigma_z\right]\psi \tag{1}$$

where the $\tau_\mu$ matrices are the Pauli matrices acting on the band index $\{d_K, d_{K'}\}$ basis, $\xi(\boldsymbol{k}) = -\boldsymbol{k}^2/2m - \mu$ is the single-particle energy without SOC, and $\lambda$ is the Ising SOC. Throughout this Supplementary Information, we will use bold symbols to denote two-dimensional vectors like the momentum $\boldsymbol{k} = (k_x, k_y)$ and arrows to denote three dimensional vectors like the spin $\vec{S} = (S_x, S_y, S_z)$. We consider the generic interaction part proposed in Ref.[8]

$$H_{\rm int} = g_5 \sum_\alpha n_{\alpha\uparrow} n_{\alpha\downarrow} + g_2 n_K n_{K'} + g_3 \sum_{\sigma\sigma'} (d^\dagger_{K\sigma} d^\dagger_{K'\sigma'} d_{K\sigma'} d_{K'\sigma}) \tag{2}$$

where $\alpha = K, K'$, $g_2$ is an interpocket density-density interaction involving the $\pm K$ pockets, $g_3$ is an interpocket pair-hopping interaction between the $K$ and $K'$ pockets, and the $g_5$ term is the intraband Hubbard repulsion. A similar model has been used in Ref.[10] in a different context. To connect with the notation in Ref.[9] we can also write this Hamiltonian in terms of spin operators explicitly as

$$H_{\rm int} = g_5 \sum_\alpha n_{\alpha\uparrow} n_{\alpha\downarrow} + (g_2 - \frac{g_3}{2}) n_K n_{K'} - g_3 \vec{S}_K \cdot \vec{S}_{K'} \tag{3}$$

where the third term becomes a spin-spin interaction. As discussed in Refs.[8,9], the microscopic origin of these couplings might be a combination of Coulomb interactions, electron-phonon interactions, or in the case of $g_3$ interactions mediated by spin fluctuations.

Next we can rewrite this interaction exactly in terms of pairing operators as

$$H_{\rm int} = (g_2 + g_3)\hat{\Delta}^\dagger_S \hat{\Delta}_S + g_5 \hat{\boldsymbol{\Delta}}^\dagger_{\boldsymbol{S'}} \cdot \hat{\boldsymbol{\Delta}}_{\boldsymbol{S'}} + (g_2 - g_3)\hat{\vec{\Delta}}^\dagger_T \cdot \hat{\vec{\Delta}}_T \tag{4}$$

where

$$\hat{\Delta}_S = \tfrac{1}{2}\psi\tau_x i\sigma_y\psi, \tag{5}$$

$$\hat{\boldsymbol{\Delta}}_{S'} = \tfrac{1}{2}\psi\{\tau_0, i\tau_z\}i\sigma_y\psi, \tag{6}$$

$$\vec{\hat{\Delta}}_T = \tfrac{1}{2}\psi i\tau_y\vec{\sigma}i\sigma_y\psi = \tfrac{1}{2}\psi i\tau_y\{-\sigma_z, i\sigma_0, \sigma_x\}\psi, \tag{7}$$

Here, $\hat{\Delta}_S$ is the spin singlet $A_1'$ channel , $\hat{\boldsymbol{\Delta}}_{S'}$ is a spin singlet $E'$ channel and $\vec{\hat{\Delta}}_T$ is the spin triplet f-wave. $\hat{\boldsymbol{\Delta}}_{S'}$ cannot lead to pairing of total zero momentum and need not be considered (in any case we assume $g_5$ to always be repulsive), so for simplicity $g_5$ is not included in the main text. We will assume that the interpocket density-density interaction is attractive, i.e., $g_2 < 0$, which favors both the s-wave and f-wave channels. We will also assume that $|g_2| > |g_3|$ so that the interaction is attractive in both channels. The choice between the two is made by the sign of $g_3$. When $g_3 = 0$ both channels have the same attraction, while the s-wave singlet ground state is favoured when $g_3 < 0$. For later convenience, we define the attraction in the singlet and triplet channels as in the main text as $g_S = g_2 + g_3$ and $g_T = g_2 - g_3$, respectively.

The inclusion of the coupling $g_5$ is useful to discuss the existence of a ferromagnetic instability within the $K$ pockets. In the Stoner picture, if we assume spins polarize in the $z$ direction and write $n_{i\uparrow} = \frac{1}{2}(n_i + M_i)$, $n_{i\downarrow} = \frac{1}{2}(n_i - M_i)$ with $i = K, K'$, where $M_i$ are the magnetizations, we can write the interaction as

$$H_{\text{int}} \approx \frac{g_5}{4}(n_K^2 + n_{K'}^2 - M_K^2 - M_{K'}^2) + (g_2 - g_3/2)n_K n_{K'} - \frac{g_3}{2}M_K M_{K'} \tag{8}$$

As usual, the Hubbard like repulsion $g_5 > 0$ favors the independent spin polarization of each band. But the existence of ferromagnetism is decided by $g_3$, where $g_3 > 0$ favors it, while $g_3 < 0$ simply favors a renormalization of the Ising SOC. Since we always assume that $g_3 < 0$, ferromagnetism is not favored.

As a final point on the origin of the different interactions, it is illustrative to consider a microscopic local model for the Coulomb repulsion, the two-orbital Slater-Kanamori Hamiltonian[11] which includes all the allowed interactions between orbital and spin degrees of freedom

$$H_{\text{int}}^{SK} = U\sum_{\gamma} n_{\gamma\uparrow}n_{\gamma\downarrow} + \frac{U'}{2}\sum_{ss',\gamma\neq\gamma'} n_{\gamma s}n_{\gamma' s'} + \sum_{ss',\gamma\neq\gamma'}\frac{J}{2}d^{\dagger}_{\gamma s}d^{\dagger}_{\gamma' s'}d_{\gamma s'}d_{\gamma' s} + \frac{J'}{2}d^{\dagger}_{\gamma s}d^{\dagger}_{\gamma s'}d_{\gamma' s'}d_{\gamma' s} \tag{9}$$

where $s =\uparrow,\downarrow$ labels the spin and $\gamma = 1, 2$ runs through the $d_i$ orbitals. $U$ is the intraorbital Hubbard repulsion, $U'$ is the interorbital Hubbard, $J$ is known as the Hund exchange coupling, and $J'$ is the pair hopping interaction. The combination of $D_{3h}$ orbital symmetry and $SU(2)$ spin rotation symmetry of the interactions force $J' = U - U' - J$, which we assume from now on.

Using the fact that the $NbSe_2$ bands have an approximately constant orbital character, to a first approximation we can perform a unitary rotation that maps $(d_1, d_2) \to (d_K, d_{K'})$ to obtain the interactions in the band basis, where we find that

$$g_2 = U - J \tag{10}$$

$$g_3 = U - U' \tag{11}$$

$$g_5 = U' + J \tag{12}$$

so that the singlet s-wave channel $\hat{\Delta}_S$ has coefficient $g_S = g_2 + g_3 = 2U - U' - J$, while the triplet f-wave interaction channel $\vec{\hat{\Delta}}_T$ has coefficient $g_T = g_2 - g_3 = U' - J$. The fact that $g_3 < 0$ favors interorbital triplet pairing is consistent with the observation made for several other multiorbital systems that the repulsive Hund coupling $J$ can lead to triplet superconductivity if it can overcome the interorbital Hubbard $U'$[12–15]. While microscopically $U' > J$, this can change when these values are renormalized as high-energy degrees of freedom are integrated out in a low-energy model.

### B. Superconducting gap equations

The next step towards modeling the collective modes in $NbSe_2$ is to solve the gap equation for the single-particle Hamiltonian 1, with local interactions given by Eq. 4 in the singlet $\hat{\Delta}_S$ and in the triplet $\vec{\hat{\Delta}}_T$ channels. We neglect the interaction in the $\hat{\boldsymbol{\Delta}}_{S'}$ channel, since it does not affect either the ground state or the collective mode to leading order.

A technical note is in order regarding SOC in this model, introduced only at the single-particle level as a k-independent constant $\lambda$. Since in 2D the DOS $\nu$ for a parabolic band of mass $m$ is constant and equal to $\nu = m/(2\pi)$, and the constant Ising SOC $\lambda$ produces just a rigid energy-shift splitting the bands but without changing their mass ($\varepsilon_\pm(\boldsymbol{k}) = \xi(\boldsymbol{k}) \pm \lambda$), then the DOS of the spin-split bands remains unchanged. Therefore, within this model the symmetry-allowed singlet-triplet mixing, which is proportional to the difference between the DOS of the spin-split bands[16], vanishes identically. We could introduce a different DOS via a k-dependent Ising SOC, which to lowest order in $\boldsymbol{k}$ would read as $-\eta \boldsymbol{k}^2/(2m)$, where the dimensionless parameter $\eta = (\nu_- - \nu_+)/(\nu_- + \nu_+)$ is the relative difference of DOS of the spin-split bands. Physically, this term arises due to the fact that the atomic SOC only affects the $d_\pm$ orbitals, whose weight in the bands is exactly 1 in the $K$ points, but it decays to about 0.8 at the Fermi level[17]. Defining the Ising SOC at the Fermi level $\lambda_F = \lambda(\boldsymbol{k} = k_F) = \lambda - \eta|\mu|$, we can thus estimate that $\eta = (\lambda - \lambda_F)/|\mu| \sim 0.2\lambda/|\mu| \sim 0.03$, which would produce a negligible singlet-triplet mixing, and therefore we do not include this term in our analysis. This conclusion has also been drawn by Ref.[8].

Interactions would also be affected by SOC in two ways. On the one hand, the triplet coupling $g_T$ would become different for the out-of-plane triplet $\hat{\Delta}_T^z = \frac{1}{2}\psi i\tau_y\sigma_x\psi$ with $A_1'$ symmetry ($g_T \to g_T^z$) and for the in-plane triplet $\hat{\boldsymbol{\Delta}}_{\boldsymbol{T}}^{\boldsymbol{xy}} = \frac{1}{2}\psi i\tau_y\{i\sigma_0, -\sigma_z\}\psi$ with $E''$ symmetry ($g_T \to g_T^{xy}$). This effect can be easily taken into account in our model just by substituting $g_T \to g_T^z$ and $g_T \to g_T^{xy}$ in the Leggett and Bardarsis-Schrieffer energy expressions given below, respectively. On the other hand, a new interaction $\left(\hat{\Delta}_S^\dagger\hat{\Delta}_T^z + \text{h.c.}\right)$ mixing the $A_1'$ singlet with the $A_1'$ out-of-plane triplet would appear, which would also induce a singlet-triplet mixed ground state. However, we will not include this term in our calculation.

The gap equation can now be derived in the following way. In the model described above, the inverse bare electron and hole Matsubara Green functions are

$$\left[G_0^{(p)}\right]^{-1}(k) = i\omega_n\tau_0\sigma_0 - H_0(\boldsymbol{k}) = i\omega_n\tau_0\sigma_0 - \left[\xi(\boldsymbol{k})\tau_0\sigma_0 + \lambda\tau_z\sigma_z\right], \tag{13}$$

$$\left[G_0^{(h)}\right]^{-1}(k) = i\omega_n\tau_0\sigma_0 + H_0^*(-\boldsymbol{k}) = i\omega_n\tau_0\sigma_0 + \left[\xi(\boldsymbol{k})\tau_0\sigma_0 + \lambda\tau_z\sigma_z\right], \tag{14}$$

and the action in imaginary time for the fermions in this model is

$$\mathcal{S}[\bar{\psi}, \psi] = -\int_k \psi^\dagger(k)\left[G_0^{(p)}\right]^{-1}(k)\psi(k) + g_S\int_x \hat{\Delta}_S^\dagger(x)\hat{\Delta}_S(x) + g_T\int_x \hat{\vec{\Delta}}_T^\dagger(x)\cdot\hat{\vec{\Delta}}_T(x), \tag{15}$$

where the pairing operators $\hat{\Delta}_S$ and $\hat{\vec{\Delta}}_T$ were defined in Eqs. 5 and 7, and we have used the shorthand notations $x \equiv (\tau, \boldsymbol{x})$, $k \equiv (i\omega_n, \boldsymbol{k})$, $\int_x \equiv \int_0^\beta d\tau \int d^2x$, and $\int_k \equiv (1/\beta)\sum_{i\omega_n}\int d^2k/(2\pi)^2$, with $\beta = (k_B T)^{-1}$ and $T$ the temperature.

Performing a Hubbard-Stratonovich transformation in the singlet $\hat{\Delta}_S$ and triplet $\hat{\vec{\Delta}}_T$ pairing channels, then going to Nambu space, and finally integrating out the fermions, we get the following effective action for the superconducting fields:

$$\mathcal{S}_{\text{eff}}[\bar{\Delta}, \Delta] = -\int_q \left[\frac{1}{g_S}\Delta_S^\dagger(q)\Delta_S(q) + \frac{1}{g_T}\vec{\Delta}_T^\dagger(q)\cdot\vec{\Delta}_T(q)\right] - \text{Tr}\left[\log G^{-1}\right], \tag{16}$$

where we have defined the inverse Bogoliubov-de Gennes (BdG) Green function as:

$$G^{-1}(k,q) = \begin{pmatrix} (2\pi)^2\beta\delta(q)[G_0^{(p)}]^{-1}(k) & -\Delta_S(-q)\tau_x i\sigma_y + \vec{\Delta}_T(-q)\cdot i\tau_y(\sigma_z, i\sigma_0, -\sigma_x) \\ \Delta_S^\dagger(q)\tau_x i\sigma_y + \vec{\Delta}_T^\dagger(q)\cdot i\tau_y(-\sigma_z, i\sigma_0, \sigma_x) & (2\pi)^2\beta\delta(q)[G_0^{(h)}]^{-1}(k) \end{pmatrix} \tag{17}$$

Let us now assume that just a single spatially-homogeneous real superconducting field $\Delta_i$ condenses. The quasi-particle energies in the superconducting state with and without SOC are therefore

$$E_\pm(\boldsymbol{k}) = \sqrt{\varepsilon_\pm^2(\boldsymbol{k}) + \Delta_i^2}, \tag{18}$$

$$E(\boldsymbol{k}) = \sqrt{\xi^2(\boldsymbol{k}) + \Delta_i^2}, \tag{19}$$

respectively, where the band energies with SOC are $\varepsilon_\pm(\boldsymbol{k}) = \xi(\boldsymbol{k}) \pm \lambda$. With this assumption, we can minimize the effective action. After performing the sum over Matsubara frequencies, we obtain the following gap equations for the

$A_1'$ singlet $\Delta_S$, the $A_1'$ triplet $\Delta_T^z$ and the $E''$ triplet $\boldsymbol{\Delta_T^{xy}}$:

$$-\frac{1}{g_S}\Delta_S = \Delta_S \int \frac{d^2k}{(2\pi)^2} \frac{\tanh\left[\frac{\beta}{2}E_+(\boldsymbol{k})\right]}{E_+(\boldsymbol{k})} + \frac{\tanh\left[\frac{\beta}{2}E_-(\boldsymbol{k})\right]}{E_-(\boldsymbol{k})}, \tag{20}$$

$$-\frac{1}{g_T}\Delta_T^z = \Delta_T^z \int \frac{d^2k}{(2\pi)^2} \frac{\tanh\left[\frac{\beta}{2}E_+(\boldsymbol{k})\right]}{E_+(\boldsymbol{k})} + \frac{\tanh\left[\frac{\beta}{2}E_-(\boldsymbol{k})\right]}{E_-(\boldsymbol{k})}, \tag{21}$$

$$-\frac{1}{g_T}\boldsymbol{\Delta_T^{xy}} = \boldsymbol{\Delta_T^{xy}} \int \frac{d^2k}{(2\pi)^2} \frac{\tanh\left[\frac{\beta}{2}(E(\boldsymbol{k})+\lambda)\right] + \tanh\left[\frac{\beta}{2}(E(\boldsymbol{k})-\lambda)\right]}{E(\boldsymbol{k})}. \tag{22}$$

As mentioned before, in general the gap equations 20 and 21 of $\Delta_S$ and $\Delta_T^z$ would be coupled, giving rise to singlet-triplet mixing. However, in the presence of just constant Ising SOC the gap equations become decoupled (as long as the chemical potential $\mu$ is much larger than the energy cutoff $\Lambda$, which is the relevant situation for $NbSe_2$, where $\mu \sim 500\text{meV} \gg \Lambda \sim \omega_D \sim 20\text{meV}$). On the other hand, the gap equation 22 for $\boldsymbol{\Delta_T^{xy}}$ shows that this in-plane triplet is suppressed by Ising SOC, and eventually killed when the Ising SOC $\lambda$ becomes bigger than the energy cutoff $\Lambda$. This is due to the fact that the $E''$ triplet $\boldsymbol{\Delta_T^{xy}}$ involves pairing between equal-spin states, but zero momentum Cooper pairs at the Fermi level can only be made with opposite spins. Furthermore, notice that the gap equations for $\Delta_S$ and $\Delta_T^z$ are identical and unaffected by the Ising SOC, which is more clearly seen after changing the momentum integration by an energy integration, $\int \frac{d^2k}{(2\pi)^2} \to \frac{N_0}{4}\int_{-\infty}^{\infty} d\varepsilon_\pm$, where $N_0 = 4\frac{m}{2\pi}$ is the total DOS in the normal state:

$$-\frac{1}{g_i}\Delta_i = N_0\Delta_i \int_{-\Lambda}^{\Lambda} d\varepsilon \frac{\tanh\left[\frac{\beta}{2}\sqrt{\varepsilon^2+\Delta_i^2}\right]}{2\sqrt{\varepsilon^2+\Delta_i^2}}, \tag{23}$$

with $i = S, T^z$. Again, this is a consequence of the fact that the $A_1'$ singlet $\Delta_S$ and the $A_1'$ triplet $\Delta_T^z$ pair electrons with opposite spins, which are those available at the Fermi level with zero momentum pairing. Rashba SOC $\propto (k_x\sigma_y - k_y\sigma_x)\tau_0$ induced by the breaking of the $xy$-plane mirror symmetry by a substrate would instead suppress the $A_1'$ singlet and the $A_1'$ triplet, while not affecting the $E''$ triplet[8]. However, we will not consider it here, leaving the study of its effect on the superconducting collective modes for future work.

It is convenient to define the dimensionless coupling constants $\bar{g}_i$ as

$$\bar{g}_i = N_0 g_i \tag{24}$$

Assuming $|g_S| > |g_T|$, the ground state is a conventional s-wave singlet superconductor with order parameter given by the usual expression at zero temperature:

$$\Delta = \frac{\Lambda}{\sinh(-1/\bar{g}_S)} \simeq 2\Lambda \exp\left(\frac{1}{\bar{g}_S}\right) \tag{25}$$

This ground state is described by the following BdG Matsubara Green function:

$$G_0^{-1}(i\omega_n, \boldsymbol{k}) = i\omega_n\tau_0\sigma_0\rho_0 - \xi(\boldsymbol{k})\tau_0\sigma_0\rho_z - \lambda\tau_z\sigma_z\rho_z + \Delta\tau_x\sigma_y\rho_y \tag{26}$$

$$\begin{aligned} G_0(i\omega_n, \boldsymbol{k}) = &\frac{1}{2}\left[\frac{i\omega_n}{(i\omega_n)^2 - E_+(\boldsymbol{k})^2} + \frac{i\omega_n}{(i\omega_n)^2 - E_-(\boldsymbol{k})^2}\right]\tau_0\sigma_0\rho_0 + \frac{1}{2}\left[\frac{\varepsilon_+(\boldsymbol{k})}{(i\omega_n)^2 - E_+(\boldsymbol{k})^2} + \frac{\varepsilon_-(\boldsymbol{k})}{(i\omega_n)^2 - E_-(\boldsymbol{k})^2}\right]\tau_0\sigma_0\rho_z + \\ &+ \frac{1}{2}\left[\frac{i\omega_n}{(i\omega_n)^2 - E_+(\boldsymbol{k})^2} - \frac{i\omega_n}{(i\omega_n)^2 - E_-(\boldsymbol{k})^2}\right]\tau_z\sigma_z\rho_0 + \frac{1}{2}\left[\frac{\varepsilon_+(\boldsymbol{k})}{(i\omega_n)^2 - E_+(\boldsymbol{k})^2} - \frac{\varepsilon_-(\boldsymbol{k})}{(i\omega_n)^2 - E_-(\boldsymbol{k})^2}\right]\tau_z\sigma_z\rho_z - \\ &- \frac{1}{2}\left[\frac{\Delta}{(i\omega_n)^2 - E_+(\boldsymbol{k})^2} + \frac{\Delta}{(i\omega_n)^2 - E_-(\boldsymbol{k})^2}\right]\tau_x\sigma_y\rho_y - \frac{1}{2}\left[\frac{\Delta}{(i\omega_n)^2 - E_+(\boldsymbol{k})^2} - \frac{\Delta}{(i\omega_n)^2 - E_-(\boldsymbol{k})^2}\right]\tau_y\sigma_x\rho_y, \end{aligned} \tag{27}$$

where $\rho_\mu$ are the Pauli matrices acting on Nambu space.

### C. Analysis of particle-particle collective modes

Particle-particle collective modes in superconductors occur generally when a second superconducting instability appears near in energy, realized as a second extremum (minimum or saddle point) of the free energy corresponding to a second solution of the gap equation. The fluctuation towards that minimum is a low energy collective mode. For example, in a two band superconductor with two $s$-wave gaps $\Delta_1$ and $\Delta_2$, the relative phase of the two gaps will be fixed by interband coupling $\lambda(\Delta_1\Delta_2^* + \Delta_1^*\Delta_2)$, but if this coupling is small compared to the intraband one, the state with $\pi$ shift in the relative phase is a saddle point, and there is a collective mode where the phases of both order parameters fluctuate out of phase ("towards" the $\pi$ shifted state). This is known as a Leggett mode[18]. A Leggett mode also occurs, as in the main text, in a one band superconductor where the Fermi surface is split by spin-orbit coupling, effectively realizing the two band scenario[19]. Similarly, if in an $s$-wave superconductor a state with different symmetry such as $p$-wave state is near in energy, the fluctuation towards the $p$ wave state is known as a Bardasis-Schrieffer mode[20]. In more complicated situations with several order parameters and spin-orbit coupling, a collective mode might fit into more than one of these categories[21].

As explained in the main text, we have two possible collective modes in our model. First, in the singlet-triplet mixed ground state, the spin-split pockets have gaps in the band basis $\Delta_\pm = \Delta_S \pm \Delta_T^z$, and the relative phase fluctuation of these two gaps is a Leggett mode. In the limit of vanishing singlet-triplet mixing, this Leggett mode becomes the fluctuation of the imaginary part of the out-of-plane $A_1'$ triplet, $\Delta_T^z = i\phi_T^z$. Second, there is a Bardarsis-Schrieffer mode consisting of the fluctuation of the imaginary part of the in-plane $E''$ triplet, $\boldsymbol{\Delta_T^{xy}} = i\boldsymbol{\phi_T^{xy}}$.

Let us assume that the interaction in the triplet channel is attractive ($g_T < 0$), so that both modes are well defined with energy $\Omega_1 < 2\Delta$ (there exists an actual pole in the propagator with vanishing imaginary part at zero temperature). To find the poles, we expand the action of Eq. 16 to quadratic order in the fluctuating fields $\Delta_T^z$ and $\boldsymbol{\Delta_T^{xy}}$. These fluctuations couple to the ground state BdG Green function via the matrices:

$$M_L = \tau_y \sigma_x \rho_x, \tag{28}$$

$$\boldsymbol{M_{BS}} = (-\tau_y \sigma_z \rho_x, \tau_y \sigma_0 \rho_y)\,. \tag{29}$$

If $\mathcal{S}_{\text{eff}}^{(0)}[\Delta]$ is the effective action of the ground state with no fluctuating fields, and $D_i^{-1}(q)$ is the inverse propagator of the Leggett ($i = L$) or Bardarsis-Schrieffer ($i = BS$) collective mode, then the total effective action expanded to quadratic order in the fluctuating fields reads as:

$$\mathcal{S}_{\text{eff}}[\Delta, \delta_i] - \mathcal{S}_{\text{eff}}^{(0)}[\Delta] = \int_q D_i^{-1}(q)\phi_i(q)\phi_i(-q). \tag{30}$$

This inverse propagator $D_i^{-1}(q)$ is related to the susceptibility $\chi_i(q)$ via:

$$D_i^{-1}(q) = -\frac{1}{g_T} + \chi_i(q), \tag{31}$$

where the susceptibility is given by:

$$\chi_i(q) = \frac{1}{2}\int_k \text{tr}[G_0(k)M_i G_0(k+q)M_i]. \tag{32}$$

Performing the traces, we obtain the following Leggett and Bardarsis-Schrieffer susceptibilities:

$$\chi_L(i\Omega_m, \boldsymbol{q}) = 2\int_k \frac{i\omega_n(i\Omega_m + i\omega_n) - [\varepsilon_+(\boldsymbol{k})\varepsilon_+(\boldsymbol{k}+\boldsymbol{q}) + \Delta^2]}{[(i\omega_n)^2 - E_+^2(\boldsymbol{k})][(i\omega_n + i\Omega_m)^2 - E_+^2(\boldsymbol{k}+\boldsymbol{q})]} + \frac{i\omega_n(i\Omega_m + i\omega_n) - [\varepsilon_-(\boldsymbol{k})\varepsilon_-(\boldsymbol{k}+\boldsymbol{q}) + \Delta^2]}{[(i\omega_n)^2 - E_-^2(\boldsymbol{k})][(i\omega_n + i\Omega_m)^2 - E_-^2(\boldsymbol{k}+\boldsymbol{q})]}, \tag{33}$$

$$\chi_{BS}(i\Omega_m, \boldsymbol{q}) = 2\int_k \frac{i\omega_n(i\Omega_m + i\omega_n) - [\varepsilon_+(\boldsymbol{k})\varepsilon_-(\boldsymbol{k}+\boldsymbol{q}) + \Delta^2]}{[(i\omega_n)^2 - E_+^2(\boldsymbol{k})][(i\omega_n + i\Omega_m)^2 - E_-^2(\boldsymbol{k}+\boldsymbol{q})]} + \frac{i\omega_n(i\Omega_m + i\omega_n) - [\varepsilon_-(\boldsymbol{k})\varepsilon_+(\boldsymbol{k}+\boldsymbol{q}) + \Delta^2]}{[(i\omega_n)^2 - E_-^2(\boldsymbol{k})][(i\omega_n + i\Omega_m)^2 - E_+^2(\boldsymbol{k}+\boldsymbol{q})]}. \tag{34}$$

Again, it is clear that the Leggett mode to the $A_1'$ triplet does not mix the bands with opposite SOC splitting, while the Bardarsis-Schrieffer mode to the $E''$ triplet does. Consequently, the constant single-particle SOC $\lambda$ will affect the Bardarsis-Schrieffer mode by increasing its energy with respect to the spinless case, while the Leggett mode energy will be independent of $\lambda$, as expected. Moreover, notice that in the limit of vanishing SOC, $\varepsilon_\pm \to \xi$, both the Leggett and the Bardarsis-Schrieffer modes become degenerate since they collapse to a single Bardarsis-Schrieffer mode to the f-wave triplet.

Let us now perform the sums over the fermionic Matsubara frequency $i\omega_n$. In the long-wavelength limit $\boldsymbol{q} \to 0$, the susceptibilities read as:

$$\chi_L(i\Omega_m) = -\int \frac{d^2k}{(2\pi)^2} \frac{\tanh\left[\frac{\beta}{2}E_+(\boldsymbol{k})\right]}{E_+(\boldsymbol{k})} \left\{1 + \frac{(i\Omega_m)^2}{[2E_+(\boldsymbol{k})]^2 - (i\Omega_m)^2}\right\} + \frac{\tanh\left[\frac{\beta}{2}E_-(\boldsymbol{k})\right]}{E_-(\boldsymbol{k})} \left\{1 + \frac{(i\Omega_m)^2}{[2E_-(\boldsymbol{k})]^2 - (i\Omega_m)^2}\right\}, \tag{35}$$

$$\chi_{BS}(i\Omega_m) = -\int \frac{d^2k}{(2\pi)^2} \frac{\tanh\left[\frac{\beta}{2}E_+(\boldsymbol{k})\right]}{E_+(\boldsymbol{k})} \{1 + B_+(i\Omega_m, \boldsymbol{k})\} + \frac{\tanh\left[\frac{\beta}{2}E_-(\boldsymbol{k})\right]}{E_-(\boldsymbol{k})} \{1 + B_-(i\Omega_m, \boldsymbol{k})\}, \tag{36}$$

where we have defined the quantity:

$$B_+(i\Omega_m, \boldsymbol{k}) = \frac{\left[\varepsilon_+(\boldsymbol{k})^2 - \varepsilon_-(\boldsymbol{k})^2 + (i\Omega_m)^2\right]\left[(\varepsilon_+(\boldsymbol{k}) - \varepsilon_-(\boldsymbol{k}))^2 - (i\Omega_m)^2\right]}{[i\Omega_m + E_+(\boldsymbol{k}) + E_-(\boldsymbol{k})][i\Omega_m + E_+(\boldsymbol{k}) - E_-(\boldsymbol{k})][i\Omega_m - E_+(\boldsymbol{k}) + E_-(\boldsymbol{k})][i\Omega_m - E_+(\boldsymbol{k}) - E_-(\boldsymbol{k})]}, \tag{37}$$

and $B_-$ equals $B_+$ when replacing $\varepsilon_\pm$ by $\varepsilon_\mp$. Notice that, in the limit of vanishing SOC, $B_\pm(i\Omega_m, \boldsymbol{k}) \to \frac{(i\Omega_m)^2}{[2E(\boldsymbol{k})]^2 - (i\Omega_m)^2}$. Using the gap equation 20 and defining the following functions:

$$F_{L\pm}(i\Omega_m) = \frac{1}{N_0}\int \frac{d^2k}{(2\pi)^2} \frac{\tanh\left[\frac{\beta}{2}E_\pm(\boldsymbol{k})\right]}{E_\pm(\boldsymbol{k})} \frac{(i\Omega_m)^2}{[2E_\pm(\boldsymbol{k})]^2 - (i\Omega_m)^2}, \tag{38}$$

$$F_{BS\pm}(i\Omega_m) = \frac{1}{N_0}\int \frac{d^2k}{(2\pi)^2} \frac{\tanh\left[\frac{\beta}{2}E_\pm(\boldsymbol{k})\right]}{E_\pm(\boldsymbol{k})} B_\pm(i\Omega_m, \boldsymbol{k}), \tag{39}$$

$$F_i(i\Omega_m) = F_{i+}(i\Omega_m) + F_{i-}(i\Omega_m), \tag{40}$$

we obtain the following collective mode propagators:

$$D_i^{-1}(i\Omega_m) = -\frac{1}{g_T} + \frac{1}{g_S} - N_0 F_i(i\Omega_m) = N_0\left\{\left(\frac{1}{|g_T|} - \frac{1}{|g_S|}\right) - F_i(i\Omega_m)\right\}, \tag{41}$$

where we have used the assumption $g_S, g_T < 0$. The collective mode energies are given by the solution $\Omega_{1i}$ of:

$$D_i^{-1}\left(i\Omega_m \to \Omega_{1i} + i0^+\right) = 0 \Rightarrow F_i\left(\Omega_{1i}\right) = \left(\frac{1}{|g_T|} - \frac{1}{|g_S|}\right). \tag{42}$$

### 1. *Collective modes at zero temperature*

Consider first the zero-temperature limit of the momentum integrations in $F_{i\pm}(\Omega)$, where $\tanh\left[\frac{\beta}{2}E_\pm(\boldsymbol{k})\right] \to 1$. Let us define the dimensionless energy of the collective mode $\bar{\Omega}_{1i}$ as its energy normalized by twice the superconducting gap, $\bar{\Omega}_{1i} = \Omega_{1i}/(2\Delta)$. When $\bar{\Omega}_{1i} < 1$ the collective mode is undamped due to the absence of a quasiparticle decay channel, while for $\bar{\Omega}_{1i} > 1$ damping becomes possible.

Let us first compute the energy of the Leggett mode. The functions $F_{L\pm}(\Omega)$ relevant for the this mode can be computed analytically by changing variables to $x = \varepsilon_\pm/\Delta$

$$F_L(\Omega) = \frac{1}{2}\int_{-\infty}^{\infty} dx \frac{1}{\sqrt{1+x^2}} \frac{\bar{\Omega}^2}{1 + x^2 - \bar{\Omega}^2} = \begin{cases} \frac{\bar{\Omega}\arcsin\left(\bar{\Omega}\right)}{\sqrt{1-\bar{\Omega}^2}} & \text{, if } |\bar{\Omega}| < 1, \\ \frac{|\bar{\Omega}|\left[-\text{argsinh}\left(\sqrt{\bar{\Omega}^2-1}\right) + i\frac{\pi}{2}\text{sign}(\bar{\Omega})\right]}{\sqrt{\bar{\Omega}^2-1}} & \text{, if } |\bar{\Omega}| > 1. \end{cases} \tag{43}$$

Assuming that the energy of the collective mode is $\Omega_{1L} < 2\Delta$, the energy of the Leggett mode is therefore the solution of:

$$\left(\frac{1}{|g_T|} - \frac{1}{|g_S|}\right) = \frac{\bar{\Omega}_{1L}\arcsin\left(\bar{\Omega}_{1L}\right)}{\sqrt{1 - \left(\bar{\Omega}_{1L}\right)^2}}. \tag{44}$$

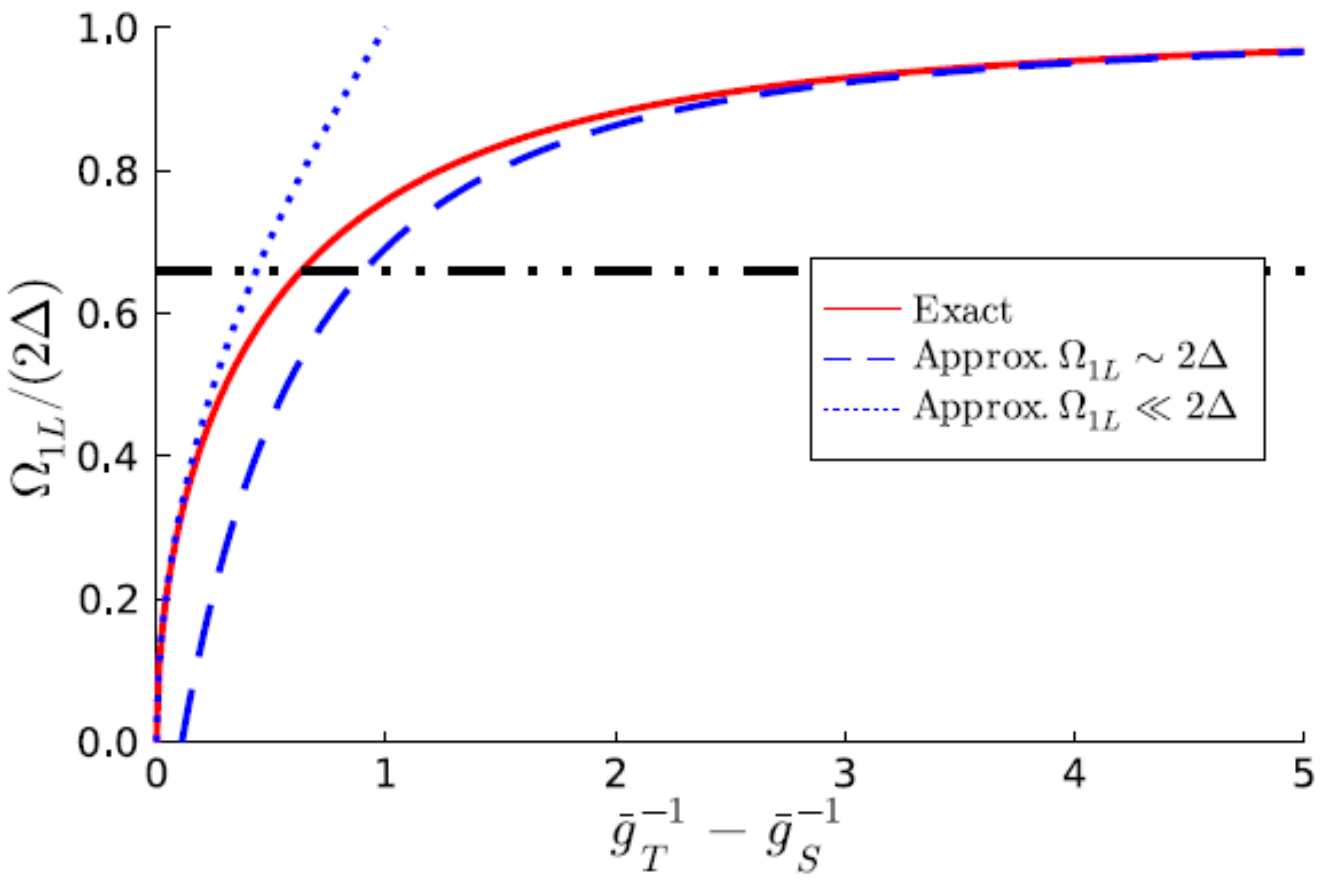

**Supplementary Figure 8:** Energy of the Leggett mode normalized by $2\Delta$ as a function of $\left(|\bar{g}_T|^{-1} - |\bar{g}_S|^{-1}\right)$. The red solid line corresponds to the exact numerical solution of Eq. 44, while the blue dotted and dashed lines are the approximate analytical expressions given in Eq. 45. We have also plotted an horizontal black dashed-dotted line at the experimental expectation value $\frac{\Omega_1}{2\Delta} = 0.66$.

This equation is reproduced in the main text where $\bar{\Omega}_{1L}$ is simply called $\bar{\Omega}_1$. The exact numerical solution for $\Omega_{1L}$ as a function of the relevant combination of the couplings $\left(|\bar{g}_T|^{-1} - |\bar{g}_S|^{-1}\right)$ is plotted in Fig. 8 together with the analytical solutions in the $|\Omega_{1L}| \ll 2\Delta$ $\left[\left(|\bar{g}_T|^{-1} - |\bar{g}_S|^{-1}\right) \ll 1\right]$ and $|\Omega_{1L}| \sim 2\Delta$ $\left[\left(|\bar{g}_T|^{-1} - |\bar{g}_S|^{-1}\right) \gg 1\right]$ limits:

$$\bar{\Omega}_{1L} \simeq \begin{cases} \sqrt{\frac{1}{|\bar{g}_T|} - \frac{1}{|\bar{g}_S|}} & \text{, if } |\bar{\Omega}_{1L}| \ll 1, \\ 1 - \frac{\pi^2}{8\left[1 + \left(|\bar{g}_T|^{-1} - |\bar{g}_S|^{-1}\right)\right]^2} & \text{, if } |\bar{\Omega}_{1L}| \sim 1. \end{cases} \tag{45}$$

On the other hand, we have not been able to obtain analytical expressions for the functions $F_{BS\pm}(\Omega)$ relevant for the Bardarsis-Schrieffer mode for a general Ising SOC $\lambda$. However, based on the suppression of the $E''$ triplet pairing by the Ising SOC (in the absence of Rashba SOC), we expect a damped Bardarsis-Schrieffer mode with energy $\Omega_{1BS} > 2\Delta$. What we can compute analytically is the series expansion of $F_{BS\pm}(\Omega)$ in small $\lambda/\Delta \ll 1$. Indeed, after the change of variables $\boldsymbol{k} \to x = \xi(\boldsymbol{k})/\Delta$, we obtain:

$$\begin{aligned} F_{BS\pm}(\Omega) &= -\frac{1}{4}\int_{-\infty}^{\infty} dx \frac{1}{\sqrt{1 + \left(x \pm \frac{\lambda}{\Delta}\right)^2}} \frac{\left(\bar{\Omega}^2 \pm x\frac{\lambda}{\Delta}\right)\left(\bar{\Omega}^2 - \frac{\lambda^2}{\Delta^2}\right)}{\bar{\Omega}^4 - \bar{\Omega}^2\left(1 + x^2 + \frac{\lambda^2}{\Delta^2}\right) + x^2\frac{\lambda^2}{\Delta^2}} = \\ &= \frac{1}{2}\left\{F_L(\Omega) + \frac{1}{2}\left(\frac{\lambda}{\Delta}\right)^2 \frac{1}{\bar{\Omega}^2 - 1}\left[1 + \frac{F_L(\Omega)}{\bar{\Omega}^2}\right] + \mathcal{O}\left[\left(\frac{\lambda}{\Delta}\right)^4\right]\right\} \end{aligned} \tag{46}$$

Despite not representing the experimentally relevant situation, this approximation predicts the expected increase of the energy of the Bardarsis-Schrieffer mode with the Ising SOC:

$$\Omega_{1BS} \simeq \begin{cases} \sqrt{(\Omega_{1L})^2 + \left(\frac{\lambda}{\Delta}\right)^2 + \mathcal{O}\left[\left(\frac{\lambda}{\Delta}\right)^4, \left(\frac{\lambda}{\Delta}\right)^2 (\Omega_{1L})^2\right]} & \text{, when } |\bar{\Omega}_{1BS}| \ll 1, \\ \Omega_{1L} + \frac{1}{2}\left(\frac{\lambda}{\Delta}\right)^2 + \mathcal{O}\left[\left(\frac{\lambda}{\Delta}\right)^3, \left(\frac{\lambda}{\Delta}\right)^2 (1 - \Omega_{1L})\right] & \text{, when } |\bar{\Omega}_{1BS}| \sim 1. \end{cases} \tag{47}$$

### *2. Collective modes at low temperature*

Let us now compute the temperature dependence of the energy of the Leggett mode in the $T \ll \Delta$ limit, with $\Delta$ the zero temperature gap. Assuming also that $T \ll \Delta(T)$, then we can make the following approximation to leading

order in temperature:

$$\tanh\left[\frac{\beta\Delta(T)}{2}\sqrt{x^2+1}\right] \simeq 1-2\exp\left(-\beta\Delta\sqrt{x^2+1}\right). \tag{48}$$

Then, the solution of the gap equation 23 gives the usual exponentially small correction to the gap at low temperatures:

$$\Delta(T) \simeq \left(1-e^{-\beta\Delta}\sqrt{\frac{2\pi}{\beta\Delta}}\right)\Delta \tag{49}$$

On the other hand, after the change of variables $x=\xi(\boldsymbol{k})/\Delta(T)$, we obtain that $F_L$ takes on the form:

$$F_L(\Omega,T) = \frac{1}{2}\int_{-\infty}^{\infty} dx \frac{1}{x^2+1-\bar{\Omega}^2}\frac{\tanh\left[\frac{\beta\Delta(T)}{2}\sqrt{x^2+1}\right]}{\sqrt{x^2+1}} \simeq F_L(\Omega) - \int_{-\infty}^{\infty} dx \frac{1}{x^2+1-\bar{\Omega}^2}\frac{\exp\left(-\beta\Delta\sqrt{x^2+1}\right)}{\sqrt{x^2+1}} \tag{50}$$

While the last integration cannot be performed analytically, we can deduce that its temperature dependence will be exponential, as in the case of the gap. Indeed, we can estimate it to be:

$$F_L(\Omega,T) \simeq F_L(\Omega)\left[1-2\gamma(\Omega)e^{-\beta\Delta}\right] \tag{51}$$

where $\gamma(\Omega)$ is a positive function of $\Omega$ of order 1. Let us mention here that $\gamma(\Omega)$ might also be slightly temperature dependent, but its temperature dependence is polynomial at most. Therefore, the inverse propagator of the Leggett mode at small $T$ gets a exponential temperature correction:

$$D_L^{-1}(\Omega) = N_0\left\{\left(\frac{1}{|\bar{g}_T|}-\frac{1}{|\bar{g}_S|}\right) - F_L(\Omega)\left[1-2\gamma(\Omega)e^{-\beta\Delta}\right]\right\} \tag{52}$$

This positive correction increases the dimensionless Leggett mode energy $\bar{\Omega}_{1L}(T)=\Omega_{1L}(T)/[2\Delta(T)]$, as expected. Indeed, since $\bar{\Omega}_{1L}(T)$ will be exponentially similar to $\bar{\Omega}_{1L}(T=0)$, to leading order in temperature we can write that $\gamma\left[\Omega_{1L}(T)\right]e^{-\beta\Delta} \simeq \gamma_L e^{-\beta\Delta}$, where we have defined $\gamma_L=\gamma\left[\Omega_{1L}(T=0)\right]\sim 1$. Therefore $\bar{\Omega}_{1L}(T)$ can be obtained from $\bar{\Omega}_{1L}(T=0)$ simply rescaling $\left(|\bar{g}_T|^{-1}-|\bar{g}_S|^{-1}\right)$ to $\left(|\bar{g}_T|^{-1}-|\bar{g}_S|^{-1}\right)\left(1+2\gamma_L e^{-\beta\Delta}\right)$, i.e.,

$$\bar{\Omega}_{1L}\left[T,\left(\frac{1}{|\bar{g}_T|}-\frac{1}{|\bar{g}_S|}\right)\right] = \bar{\Omega}_{1L}\left[T=0,\left(\frac{1}{|\bar{g}_T|}-\frac{1}{|\bar{g}_S|}\right)\left(1+2\gamma_L e^{-\beta\Delta}\right)\right]. \tag{53}$$

The same argument applies to the energy of the Bardarsis-Schrieffer mode, and therefore its normalized energy is also exponentially increased by temperature in the $T \ll \Delta$ limit. This increase of the normalized energy with temperature is expected, since the collective mode should be overdamped (i.e., $\Omega_1 \geq 2\Delta$) at the critical temperature[22].

### D. Computation of the tunneling current

Most works to date have attributed satellite peaks to particle-hole excitations like spin-waves, which become undamped in the presence of superconductivity. However, particle-particle collective modes should also leave fingerprints in the spectral function. We now present a simplified calculation along the lines of Ref.[23] to show how coupling to a particle-particle collective mode leads to peaks in the STM tunneling current.

In the presence of bosonic excitations, the tunneling spectra has both an elastic contribution due to the boson contribution to the fermion self-energy as well as an inelastic contribution due to the emission of real bosons in the tunneling process. The inelastic contribution is sizable when tunneling is effective at all momenta, since there is a large phase space for the boson in tunneling[24,25]. Conversely, the inelastic tunneling contribution can be neglected when tunneling is dominated by small momenta. We will not consider the inelastic contribution in this work.

We can now estimate the elastic contribution to the tunneling DOS due to the Leggett and Bardarsis-Schrieffer modes in the following way. The bare BdG Matsubara Green function unrenormalized by the boson is given in Eq. 27. On the other hand, in the previous subsection we have obtained the collective mode propagator at zero momentum to be:

$$D_i^{-1}(i\Omega_m) = N_0\left[\left(\frac{1}{|\bar{g}_T|}-\frac{1}{|\bar{g}_S|}\right) - F_i(i\Omega_m)\right] = N_0\left[F_i\left(\Omega_{1i}\right)-F_i(i\Omega_m)\right] \tag{54}$$

where we have used the fact that $D_i^{-1}(i\Omega_m \to \Omega_{1i} + i0^+) = 0$. In order to simplify the calculations, we expand the propagator about its poles at $i\Omega = \pm\Omega_{1i}$ to leading order:

$$D_i(i\Omega_m) \simeq \frac{Z_i}{(\Omega_{1i})^2 - (i\Omega_m)^2} \tag{55}$$

where we have defined the residues of the Leggett and Bardarsis-Schrieffer propagators

$$Z_L = \frac{8\Delta^2}{N_0} \frac{1 - \left(\bar{\Omega}_{1L}\right)^2}{1 + \frac{(|\bar{g}_T|^{-1} - |\bar{g}_S|^{-1})}{\left(\bar{\Omega}_{1L}\right)^2}}, \tag{56}$$

$$Z_{BS} \simeq \frac{8\Delta^2}{N_0} \left\{ \frac{1 - \left(\bar{\Omega}_{1BS}\right)^2}{1 + \frac{(|\bar{g}_T|^{-1} - |\bar{g}_S|^{-1})}{\left(\bar{\Omega}_{1BS}\right)^2}} + \left[ 1 - \frac{\left(|\bar{g}_T|^{-1} - |\bar{g}_S|^{-1}\right)\left[1 + \left(|\bar{g}_T|^{-1} - |\bar{g}_S|^{-1}\right)\right]}{\left[\left(|\bar{g}_T|^{-1} - |\bar{g}_S|^{-1}\right) + \left(\bar{\Omega}_{1BS}\right)^2\right]^2} \right] \left(\frac{\lambda}{\Delta}\right)^2 + \mathcal{O}\left[\left(\frac{\lambda}{\Delta}\right)^4\right] \right\}, \tag{57}$$

respectively. We observe that the residue of the propagator scales with $\Delta^2$, reflecting the weight of these bosons are supressed as $\Delta \to 0$.

The action coupling the fermions $\Psi$ to the collective boson $\phi_i$ reads, in Nambu space:

$$\mathcal{S} = -\frac{1}{2} \int_k \Psi^\dagger(k) G_0^{-1}(k) \Psi(k) + \int_q \phi_i(q) D^{-1}(q) \phi_i(-q) - \frac{\alpha_i}{\sqrt{2}} \int_k \int_q \Psi^\dagger(k) \phi_i(q) M_i \Psi(k+q) \tag{58}$$

where the matrices $M_i$ via which the Leggett and Bardarsis-Schrieffer modes couple to the fermions were defined in Eqs. 28-29, and $\alpha_i$ are corresponding coupling strengths.

The coupling to the collective mode induces a quasiparticle self-energy[26] $\Xi(i\omega_n)$, which consists of a normal part $\tilde{\Sigma}(i\omega_n)$ in the $\rho_0$ channel including lifetime effects and an anomalous part $\tilde{\Phi}(i\omega_n)$ in the $\rho_y$ channel modeling the effect of superconducting pairing. To one loop, the self-energy reads:

$\Xi_i(i\omega_n) =$ 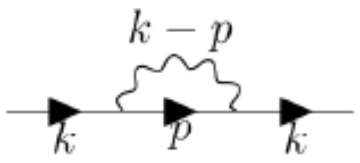

$$= \alpha_i^2 \int_p M_i G_0(p) M_i D_i(k-p) \equiv \alpha_i^2 \frac{1}{\beta} \sum_{i\nu_m} \int \frac{d^2p}{(2\pi)^2} M_i G_0(i\nu_m, \boldsymbol{p}) M_i D_i(i\omega_n - i\nu_m) \tag{59}$$

As we will see below, we can parametrize this self-energy as:

$$\Xi_i(i\omega_n) = \tilde{\Sigma}_i(i\omega_n)\tau_0\sigma_0\rho_0 + \tilde{\Phi}_i(i\omega_n)\tau_x\sigma_y\rho_y \tag{60}$$

Therefore, the renormalized inverse Green function becomes:

$$G_i^{-1}(i\omega_n, \boldsymbol{k}) = G_0^{-1}(i\omega_n, \boldsymbol{k}) + \Xi_i(i\omega_n) = \Sigma_i(i\omega_n)\tau_0\sigma_0\rho_0 - \xi(\boldsymbol{k})\tau_0\sigma_0\rho_z - \lambda\tau_z\sigma_z\rho_z + \Phi_i(i\omega_n)\tau_x\sigma_y\rho_y \tag{61}$$

where, for convenience, we have redefined $\Sigma_i$ and $\Phi_i$ by introducing the frequency $i\omega_n$ and the gap $\Delta$, respectively, i.e.:

$$\Sigma_i(i\omega_n) = i\omega_n + \tilde{\Sigma}_i(i\omega_n), \tag{62}$$

$$\Phi_i(i\omega_n) = \Delta + \tilde{\Phi}_i(i\omega_n). \tag{63}$$

Thus, we arrive at the following renormalized Matsubara Green function:

$$\begin{aligned}
G_i(i\omega_n,\boldsymbol{k}) = &\frac{1}{2}\left[\frac{\Sigma_i(i\omega_n)}{\Sigma_i(i\omega_n)^2-\Phi_i(i\omega_n)^2-\varepsilon_+(\boldsymbol{k})^2}+\frac{\Sigma_i(i\omega_n)}{\Sigma_i(i\omega_n)^2-\Phi_i(i\omega_n)^2-\varepsilon_-(\boldsymbol{k})^2}\right]\tau_0\sigma_0\rho_0+\\
&+\frac{1}{2}\left[\frac{\varepsilon_+(\boldsymbol{k})}{\Sigma_i(i\omega_n)^2-\Phi_i(i\omega_n)^2-\varepsilon_+(\boldsymbol{k})^2}+\frac{\varepsilon_-(\boldsymbol{k})}{\Sigma_i(i\omega_n)^2-\Phi_i(i\omega_n)^2-\varepsilon_-(\boldsymbol{k})^2}\right]\tau_0\sigma_0\rho_z+\\
&+\frac{1}{2}\left[\frac{\Sigma_i(i\omega_n)}{\Sigma_i(i\omega_n)^2-\Phi_i(i\omega_n)^2-\varepsilon_+(\boldsymbol{k})^2}-\frac{\Sigma_i(i\omega_n)}{\Sigma_i(i\omega_n)^2-\Phi_i(i\omega_n)^2-\varepsilon_-(\boldsymbol{k})^2}\right]\tau_z\sigma_z\rho_0+\\
&+\frac{1}{2}\left[\frac{\varepsilon_+(\boldsymbol{k})}{\Sigma_i(i\omega_n)^2-\Phi_i(i\omega_n)^2-\varepsilon_+(\boldsymbol{k})^2}-\frac{\varepsilon_-(\boldsymbol{k})}{\Sigma_i(i\omega_n)^2-\Phi_i(i\omega_n)^2-\varepsilon_-(\boldsymbol{k})^2}\right]\tau_z\sigma_z\rho_z-\\
&-\frac{1}{2}\left[\frac{\Phi_i(i\omega_n)}{\Sigma_i(i\omega_n)^2-\Phi_i(i\omega_n)^2-\varepsilon_+(\boldsymbol{k})^2}+\frac{\Phi_i(i\omega_n)}{\Sigma_i(i\omega_n)^2-\Phi_i(i\omega_n)^2-\varepsilon_-(\boldsymbol{k})^2}\right]\tau_x\sigma_y\rho_y-\\
&-\frac{1}{2}\left[\frac{\Phi_i(i\omega_n)}{\Sigma_i(i\omega_n)^2-\Phi_i(i\omega_n)^2-\varepsilon_+(\boldsymbol{k})^2}-\frac{\Phi_i(i\omega_n)}{\Sigma_i(i\omega_n)^2-\Phi_i(i\omega_n)^2-\varepsilon_-(\boldsymbol{k})^2}\right]\tau_y\sigma_x\rho_y
\end{aligned} \tag{64}$$

To obtain the tunneling DOS $N_i(\omega)$ we first define the integration over momenta of the spectral function

$$\rho_i(\omega) = -\frac{1}{\pi}\int\frac{d^2k}{(2\pi)^2}G_i''(i\omega_n\to\omega+i0^+,\boldsymbol{k}) \tag{65}$$

where $f''(x) = \text{Im}\,[f(x)]$. $N_i(\omega)$ is then obtained from the upper left block $(\rho_0+\rho_z)$ as

$$N_i(\omega) = \text{tr}\left[\left(\frac{\rho_0+\rho_z}{2}\right)\rho_i(\omega)\right] \tag{66}$$

Transforming the momentum integration into an energy integration, $\int\frac{d^2k}{(2\pi)^2}\to\frac{N_0}{4}\int_{-\infty}^{\infty}d\varepsilon_\pm$, we obtain

$$N_i(\omega) = N_0\text{Im}\left[\frac{\Sigma_i(\omega)}{\sqrt{\Phi_i(\omega)^2-\Sigma_i(\omega)^2}}\right] \tag{67}$$

In the bare case without the boson renormalization ($\Sigma_i(\omega)=\omega+i\delta$, $\Phi_i(\omega)=\Delta$) we find:

$$\rho_0(\omega) = \frac{N_0}{4}\left[\frac{|\omega|}{\sqrt{\omega^2-\Delta^2}}\tau_0\sigma_0\rho_0-\frac{\Delta\text{sign}(\omega)}{\sqrt{\omega^2-\Delta^2}}\tau_x\sigma_y\rho_y\right]\Theta(\omega^2-\Delta^2), \tag{68}$$

where $\Theta(x)$ is the Heaviside step function, and thus we obtain

$$N_0(\omega) = N_0\frac{|\omega|}{\sqrt{\omega^2-\Delta^2}}\Theta(\omega^2-\Delta^2), \tag{69}$$

so the DOS is zero for $\omega<\Delta$, displays the usual square-root-singular coherence peak at $\omega=\Delta$, and decays to the normal DOS at large $\omega$.

Extra structure in the $\omega$ dependence of the DOS can appear when there is structure in $\Sigma_i(\omega)$ and $\Phi_i(\omega)$. In the presence of a boson with gap $\Omega_{1i}$, both self energies develop a singularity at $\omega=\Delta+\Omega_{1i}$, when fermionic quasiparticles can decay into the boson. Let us therefore compute the self-energy to one loop in order to derive the renormalized DOS. Applying the spectral representation to the Matsubara Green functions of both electrons and bosons, and performing the sum over Matsubara frequencies using the standard procedure, the imaginary part of the self-energy can be written as:

$$\Xi_i''(\omega) = \alpha_i^2\int_{-\infty}^{\infty}d\epsilon M_i\rho_0(\epsilon)M_iD_i''(\omega-\epsilon)\left[1+n_B(\omega-\epsilon)-n_F(\epsilon)\right], \tag{70}$$

where $n_B$ and $n_F$ are the Bose-Einstein and Fermi-Dirac distribution functions, respectively.

Now, substituting the particular form of the approximate boson propagator that we are considering, $D_i''(\epsilon)$ =

$\frac{\pi Z_i}{2\Omega_{1i}}\left[\delta\left(\epsilon-\Omega_{1i}\right)-\delta\left(\epsilon+\Omega_{1i}\right)\right]$, the imaginary part of the self-energy becomes:

$$\Xi_i''(\omega)=\frac{\pi\alpha_i^2 Z_i}{2\Omega_{1i}}M_i\left\{\rho_0\left(\omega-\Omega_{1i}\right)\left[n_B\left(\Omega_{1i}\right)+1-n_F\left(\omega-\Omega_{1i}\right)\right]+\rho_0\left(\omega+\Omega_{1i}\right)\left[n_B\left(\Omega_{1i}\right)+n_F\left(\omega+\Omega_{1i}\right)\right]\right\}M_i. \quad (71)$$

At zero temperature, the previous expression simplifies to:

$$\Xi_i''(\omega)=\frac{\pi\alpha_i^2 Z_i}{2\Omega_{1i}}M_i\left\{\rho_0\left(\omega-\Omega_{1i}\right)\Theta\left(\omega-\Omega_{1i}\right)+\rho_0\left(\omega+\Omega_{1i}\right)\Theta\left(-\omega-\Omega_{1i}\right)\right\}M_i \quad (72)$$

and we see that in this approximation the self-energy simply gets at copy of $\rho_0$ shifted by $\Omega_{1i}$. For $\omega>0$, this further reduces to:

$$\Xi_i''(\omega>0)=\frac{\pi\alpha_i^2 Z_i N_0}{8\Omega_{1i}}\left[\frac{|\omega-\Omega_{1i}|}{\sqrt{\left(\omega-\Omega_{1i}\right)^2-\Delta^2}}\tau_0\sigma_0\rho_0+\frac{\Delta}{\sqrt{\left(\omega-\Omega_{1i}\right)^2-\Delta^2}}\tau_x\sigma_y\rho_y\right]\Theta\left(\omega-\Delta-\Omega_{1i}\right). \quad (73)$$

Therefore, the zero-temperature imaginary part of the normal and anomalous self-energies read as:

$$\tilde{\Sigma}_i''(\omega>0)=\frac{\pi\alpha_i^2 Z_i N_0}{8\Omega_{1i}}\frac{|\omega-\Omega_{1i}|}{\sqrt{\left(\omega-\Omega_{1i}\right)^2-\Delta^2}}\Theta\left(\omega-\Delta-\Omega_{1i}\right), \quad (74)$$

$$\tilde{\Phi}_i''(\omega>0)=\frac{\pi\alpha_i^2 Z_i N_0}{8\Omega_{1i}}\frac{\Delta}{\sqrt{\left(\omega-\Omega_{1i}\right)^2-\Delta^2}}\Theta\left(\omega-\Delta-\Omega_{1i}\right). \quad (75)$$

The nonzero imaginary part of the self-energy means that there is a finite lifetime due to the coupling to the boson. In particular, the imaginary part of both the normal $\tilde{\Sigma}_i''(\omega)$ and the anomalous $\tilde{\Phi}_i''(\omega)$ self-energies have a square-root singularity at $|\omega|=\Delta+\Omega_1$, and are zero below it. By Kramers-Kronig, the real parts must also display the same singularity.

We can then go back to Eq. 67 to compute the DOS, and the peaks in the self-energies will lead to a peak in the DOS at $\Delta+\Omega_{1i}$. Since, despite the different matrix structure of our problem, the analytical expressions for the self-energy we have obtained are exactly the same as in Ref.[23], the functional form of $N(\omega)$ will be that of Fig. 13(a) of Ref.[23], displaying the mentioned peak at $\Delta+\Omega_{1i}$. When computing the self-energy to higher orders, further satellites at $|\omega|=\Delta+n\Omega_{1i}$ are expected to appear in the DOS. The existence of higher harmonics has not been commonly reported in strongly correlated systems, although phonon harmonics have been observed in Pb, see, for instance see S. Fig. 6.

---